\documentclass[PAPER, atlasdraft=false,cernpreprint, UKenglish, texmf, orcidlogo]{atlasdoc}
\usepackage{atlaspackage}
\usepackage{multirow}
\usepackage{booktabs}
\usepackage{atlasbiblatex}

\usepackage{siunitx} %
\usepackage{array}
\newcolumntype{C}[1]{>{\centering\arraybackslash}p{#1}}

\usepackage{atlasphysics}
\graphicspath{{logos/}{figures/}}

\usepackage{ANA-HIGP-2025-10-PAPER-defs}

\AtlasTitle{Study of Higgs boson pair production in the $HH \rightarrow b  \overline{b} \gamma \gamma$ final state with 308 fb$^{-1}$ of  data collected at $\sqrt{s} =$ 13~TeV and 13.6~TeV by the ATLAS experiment}
\AtlasAbstract{
A search for Higgs boson pair production in the $b \bar{b} \gamma \gamma$ final state is performed.
The proton--proton collision dataset corresponds to an integrated luminosity of 308 fb$^{-1}$, consisting of two samples, 140 fb$^{-1}$ at a centre-of-mass energy of  $\sqrt{s} = 13\,\text{TeV}$ and 168 fb$^{-1}$ at  $\sqrt{s} =13.6\,\text{TeV}$, recorded between 2015 and 2024 by the ATLAS detector at the CERN Large Hadron Collider.
In addition to a larger dataset, this analysis improves upon the previous search in the same final state through several methodological and technical developments.
The Higgs boson pair production cross section divided by the Standard Model prediction is found to be  $\mu_{HH} =
0.9^{+1.4}_{-1.1}$ ($\mu_{HH} = 1^{+1.3}_{-1.0}$ expected), which translates into a 95\% confidence-level upper limit of $\mu_{HH}<3.7$.
At the same confidence level the Higgs self-coupling modifier is constrained to be in the range $-1.6 < \kappa_\lambda < 6.6$  ($-1.8 < \kappa_\lambda <  6.9$ expected).
}

\AtlasRefCode{HIGP-2025-10}

\PreprintIdNumber{CERN-EP-2025-140}

\AtlasJournalRef{Phys.\ Lett.\ B. 876 (2026) 140280}
\AtlasDOI{https://doi.org/10.1016/j.physletb.2026.140280}

\AtlasCoverEgroupAnalysisTeam{atlas-higp-2025-10-analysis-team@cern.ch}

\hypersetup{pdftitle={ATLAS document},pdfauthor={The ATLAS Collaboration}}

\begin{document}

\maketitle

\section{Introduction}
\label{sec:intro}
%

\begin{figure}%
\centering
\subfloat[]{
\label{fig:feyn_ggf_tri}
\includegraphics[width=0.43\textwidth]{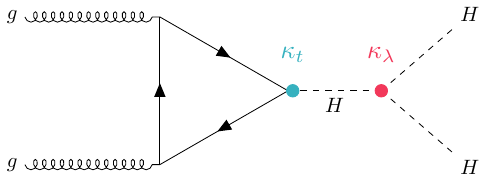} }
\quad\quad
\subfloat[]{
\label{fig:feyn_ggf_box}
\includegraphics[width=0.43\textwidth]{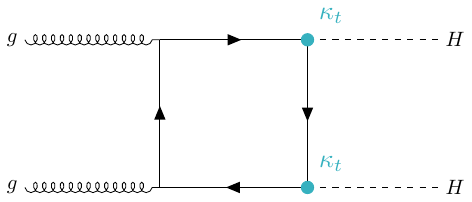} }\\

\centering
\subfloat[]{
\label{fig:feyn_VBF_tri}
\includegraphics[width=0.25\textwidth]{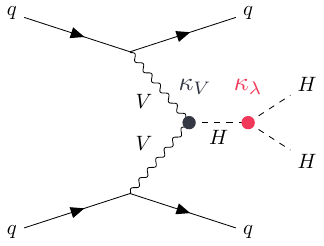} }
\quad\quad
\subfloat[]{
\label{fig:feyn_VBF_double}
\includegraphics[width=0.25\textwidth]{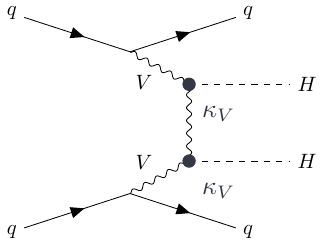} }
\quad\quad
\subfloat[]{
\label{fig:feyn_VBF_HHVV}
\includegraphics[width=0.25\textwidth]{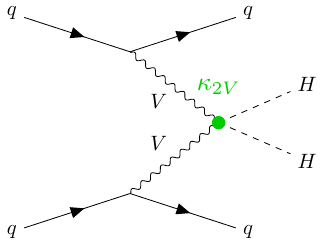} }\\

\caption[Feynman diagrams of $HH$ boson production.]{Leading-order Feynman
diagrams for (a,b) the dominant ggF production and (c-e) VBF production
of Higgs boson pairs, where (a,c) involve the trilinear Higgs coupling,
(b,d) have two independent single Higgs vertices and (e) contains the
$VVHH$ vertex.  The coupling modifiers for the $ttH$, $VVH$ and $VVHH$
couplings are represented by $\kappa_t$, $\kappa_V$, $\kappa_{2V}$,
respectively. }
\end{figure}

Following the discovery of the Higgs boson ($H$)~\cite{Englert:1964et,
Higgs:1964ia,Higgs:1964pj,Guralnik:1964eu,Higgs:1966ev,Kibble:1967sv} by
the ATLAS~\cite{HIGG-2012-27} and CMS~\cite{CMS-HIG-12-028} Collaborations,
a central objective of the Large Hadron Collider (LHC) \cite{Evans:2008zzb}
programme is the exploration of the Higgs
potential, which governs electroweak symmetry breaking (EWSB).
In the Standard Model (SM) this potential is fully determined by
the measured vacuum expectation value, $v\approx 246\, \GeV$, and the Higgs
boson mass, $m_H \approx 125 \, \GeV$.
Deviations from the predicted form would indicate new physics with profound implications for particle physics and cosmology~\cite{Horn:2020wif,Reichert:2017puo,Noble:2007kk,Markkanen:2018pdo}.
After EWSB, the Higgs potential is given by
\[
V(H) = \frac{1}{2}m_H^2 H^2 + \lambda_3 vH^3 + \frac{1}{4}\lambda_4 H^4,
\]
with $\lambda_3 = \lambda_4 = \lambda_{\mathrm{SM}} = \frac{m_H^2}{2v^2}$
in the SM\@.  At the LHC it can be probed by studying the trilinear Higgs coupling
which is accessible via Higgs boson pair ($HH$) production.  The trilinear
Higgs coupling is parameterised by a coupling modifier
$\kappa_\lambda = \frac{\lambda_{\text{3,meas}}}{\lambda_{\text{SM}}}$,
where $\lambda_{\text{3,meas}}$ denotes its measured value.
The quartic Higgs coupling, $\lambda_4$,  for which the experimental
sensitivity is very limited, has been marginally constrained so far by
a search for triple Higgs boson production~\cite{HIGP-2024-32}.
In gluon--gluon fusion (ggF), Higgs boson pairs are produced through diagrams involving either the trilinear Higgs vertex
(Figure~\ref{fig:feyn_ggf_tri})
or two top-quark Yukawa interaction vertices
(Figure~\ref{fig:feyn_ggf_box}).
Vector-boson fusion (VBF) production of Higgs boson pairs receives
contributions from the trilinear Higgs vertex
(Figure~\ref{fig:feyn_VBF_tri}),
$VVH$ vertices
(Figure~\ref{fig:feyn_VBF_double}),
as well as the quartic $VVHH$ vertex
(Figure~\ref{fig:feyn_VBF_HHVV}),
which exhibits sensitivity to non-SM
contributions primarily at large values of the $HH$ invariant mass. In both production modes the
diagrams interfere destructively and the sensitivity to $\kappa_\lambda$
lies mainly close to the $HH$ production threshold.  The production cross
section depends in addition on the top-quark Yukawa coupling in the
case of ggF, and on the $VVH$ ($\kappa_V$) and $VVHH$ ($\kappa_{2V}$)
coupling modifiers for VBF.

The predicted cross sections in the SM are very small, namely
$\sigma^{HH}_{\mathrm{ggF}} =
30.8^{+2.0}_{-7.1}$~fb~\cite{Grazzini:2018bsd, Dawson:1998py,
Borowka:2016ehy, Baglio:2018lrj, deFlorian:2013jea, Shao:2013bz,
deFlorian:2015moa, Baglio:2020wgt,Heinrich:2022idm,Bagnaschi:2023rbx} and
\mbox{$\sigma^{HH}_{\mathrm{VBF}}=1.69\pm
0.05$~fb}~\cite{Baglio:2013,Ling:2014,Dreyer:2018qbw,Dreyer:2020xaj}
for proton--proton ($pp$) collisions at a centre-of-mass energy of $\sqrt{s}=13$~\TeV,\footnote{
The cross section
prediction changed slightly compared to the one used in the previous
iteration of this analysis as detailed in Ref.~\cite{HDBS-2021-10} due to a
new set of parton distribution functions and the inclusion of electroweak
corrections for VBF.
} about three orders of magnitude smaller than for
single Higgs boson production.  At 13.6~\TeV they rise to
$34.1^{+2.2}_{-7.9}$~fb and $1.87\pm 0.05$~fb, respectively~\cite{
Grazzini:2018bsd, Dawson:1998py,
Borowka:2016ehy, Baglio:2018lrj, deFlorian:2013jea, Shao:2013bz,
deFlorian:2015moa, Baglio:2020wgt,Heinrich:2022idm,Bagnaschi:2023rbx,Baglio:2013,Ling:2014,Dreyer:2018qbw,Dreyer:2020xaj}.

This Letter reports on an updated analysis of $HH$ production in the
$HH \rightarrow b\bar{b} \gamma\gamma$ channel. Although this channel
represents only 0.26\% of the $HH$ events, it benefits from the excellent
di-photon mass resolution and offers the highest efficiency among the
leading search channels near the $HH$ production threshold, where the
sensitivity to $\kappa_\lambda$ is the greatest.  The analysis uses the full
LHC Run-2 $pp$ collision dataset of $140\, \ifb$ at $\sqrt{s}=13$~\TeV
collected by the ATLAS experiment and a portion of the Run-3 dataset
(2022--2024) corresponding to $168 \, \ifb$ at $\sqrt{s}=13.6$~\TeV.
After pre-selection, events are categorised
according to  the $b\bar{b}\gamma\gamma$ invariant mass, signal purity, and data-taking period.
A fit to the di-photon mass distribution is then performed across categories, with the background modelled using analytic functions.
The analysis is improved in several aspects compared to
the previous iteration~\cite{HDBS-2021-10}. The most significant are a new $b$-tagging
algorithm based on transformer neural
networks (GN2)~\cite{FTAG-2023-05,FTAG-2019-02,FTAG-2020-08,FTAG-2018-01}
and a kinematic fit to improve the $b\bar{b}$ and \bbyy
invariant mass resolutions.

During LHC Run 2, an extensive programme was carried out by ATLAS and CMS to
search for $HH$ production and constrain $\kappa_\lambda$ in the
$b\bar{b} \gamma\gamma$ channel~\cite{HDBS-2021-10,CMS-HIG-19-018}, as in several other final states~\cite{HDBS-2019-29,  HDBS-2022-02,HDBS-2019-27,HDBS-2019-04,HDBS-2019-02,CMS-HIG-20-010,CMS-HIG-20-005,CMS-HIG-22-006,CMS-HIG-21-002,CMS-HIG-21-005}. Combining
all channels, ATLAS and CMS constrained at 95\% confidence
level (CL) the $HH$ signal strength $\mu_{HH}$,
defined as the observed Higgs boson pair production cross section divided
by the SM prediction, to be $\mu_{HH} < 2.9$~\cite{HDBS-2021-18} and
$\mu_{HH} < 3.4$~\cite{CMS-HIG-22-001} respectively, assuming the SM $HH$ kinematics.
The trilinear Higgs coupling modifier was constrained at the same confidence level
by ATLAS to be $-1.2 < \kappa_\lambda < 7.2$~\cite{HDBS-2021-18} and by CMS
to be $-1.4 < \kappa_\lambda < 7.8$~\cite{CMS-HIG-23-006}.
%
%

%
%
%
%
%
%
%


%
\section{ATLAS detector}
\label{sec:detector}
%

\newcommand{\AtlasCoordFootnote}{%
ATLAS uses a right-handed coordinate system with its origin at the nominal interaction point (IP)
in the centre of the detector and the \(z\)-axis along the beam pipe.
The \(x\)-axis points from the IP to the centre of the LHC ring,
and the \(y\)-axis points upwards.
Polar coordinates \((r,\phi)\) are used in the transverse plane,
\(\phi\) being the azimuthal angle around the \(z\)-axis.
The pseudorapidity is defined in terms of the polar angle \(\theta\) as \(\eta = -\ln \tan(\theta/2)\) and is equal to the rapidity
$ y = \frac{1}{2} \ln \left( \frac{E + p_z}{E - p_z} \right) $ in the relativistic limit.
Angular distance is measured in units of \(\Delta R \equiv \sqrt{(\Delta
y)^{2} + (\Delta\phi)^{2}}\).}

The ATLAS experiment~\cite{PERF-2007-01,GENR-2019-02} at the LHC is a multipurpose particle detector
with a forward--backward symmetric cylindrical geometry and a near \(4\pi\) coverage in
solid angle\footnote{\AtlasCoordFootnote}.
It consists of an inner tracking detector surrounded by a thin superconducting solenoid
providing a \qty{2}{\tesla} axial magnetic field, electromagnetic and hadronic calorimeters, and a muon spectrometer.
The inner tracking detector covers the pseudorapidity range \(|\eta| < 2.5\).
It consists of silicon pixel, silicon microstrip, and transition radiation tracking detectors.
Lead/liquid-argon (LAr) sampling calorimeters provide electromagnetic (EM) energy measurements
with high granularity within the region \(|\eta|< 3.2\).
A steel/scintillator-tile hadronic calorimeter covers the central pseudorapidity range (\(|\eta| < 1.7\)).
The endcap and forward regions are instrumented with LAr calorimeters
for EM and hadronic energy measurements up to \(|\eta| = 4.9\).
The muon spectrometer surrounds the calorimeters and is based on
three large superconducting air-core toroidal magnets with eight coils each.
The field integral of the toroids ranges between \num{2.0} and \qty{6.0}{\tesla\metre}
across most of the detector.
The muon spectrometer includes a system of precision tracking chambers up to \(|\eta| = 2.7\) and fast detectors for triggering up to \(|\eta| = 2.4\).
The luminosity is measured mainly by the LUCID--2 detector which is located close to the beampipe.
A two-level trigger system was used to select events~\cite{TRIG-2016-01,TRIG-2022-01}.
The first-level trigger is implemented in hardware and it uses a subset of the detector information
to accept events at a rate close to \qty{100}{\kHz}.
This is followed by a software-based trigger that
reduced the accepted rate of complete events to \qty{1.25}{\kHz} and \qty{3}{\kHz} on average in Run~2 and Run~3, respectively,
depending on the data-taking conditions.
A software suite~\cite{SOFT-2022-02} is used in data simulation, in the reconstruction
and analysis of real and simulated data, in detector operations, and in the trigger and data acquisition
systems of the experiment.


%
\section{Data and simulation samples}
\label{sec:dataMC}
%

The analysis presented in this Letter uses $pp$-collision data collected by the ATLAS experiment during the LHC Run~2 at $\sqrt s =13$~\TeV (from 2015 to 2018) and during the first part of the LHC Run~3 at $\sqrt s =13.6$~\TeV (from 2022 to 2024).
Data quality requirements for Run~2 and Run~3 are detailed in Ref.~\cite{DAPR-2018-01}.
The resulting integrated luminosity is $140.1 \pm 1.2$~fb$^{-1}$
for the Run-2 dataset~\cite{DAPR-2021-01} and $168.0 \pm 6.7$~fb$^{-1}$
for the Run-3 dataset,\footnote{
The preliminary luminosity for the 2024 data was estimated by using the number of $Z \to ee$ and $Z \to \mu\mu$ events relative to those in the 2022/2023 data within a preliminary uncertainty of 5\%. Combined with the 2\% uncertainty for 2022/2023~\cite{ATL-DAPR-PUB-2024-001}, this results in a luminosity-weighted uncertainty for the Run-3 dataset of 4\%.} respectively.

This analysis searches for Higgs boson pair production in the $b\bar{b}\gamma\gamma$ final state, considering as signal both the dominant ggF and the subdominant VBF production modes.
For the Monte Carlo (MC) simulation of the ggF signal, events are generated using \POWHEGBOX[v2]~\cite{Nason:2004rx, Frixione:2007vw, Alioli:2010xd,powheg-sm,Heinrich:2019bkc} with parton distribution functions (PDFs) from the \PDFforLHC[21]~\cite{Butterworth:2015oua} set.
In addition to the SM sample, where the value of $\kappa_\lambda$ is set to 1, other samples corresponding to beyond-the-SM (BSM) scenarios are generated with $\kappa_\lambda=-1$, $0$, $2.5$, $5$ and $10$, respectively.
For VBF $HH$ signal events, \MGNLO~\cite{Alwall:2014hca} is used in the leading-order (LO) generation with the five-flavour scheme and the \NNPDF[3.0nlo]~\cite{Ball:2014uwa} PDF.
Furthermore, 12 additional VBF $HH$ samples are generated with the same settings as the nominal sample, but using BSM combinations of the coupling modifiers \kl, \kvv and \kv.
In both cases, \PYTHIA[8.309]~\cite{Bierlich:2022pfr} is used for the simulation of the Higgs boson decays, parton showering, hadronisation and underlying event with the A14 set of tune parameters~\cite{ATL-PHYS-PUB-2014-021}.
In the generation of the Higgs boson decay into photons~\cite{Nason:2004rx, Frixione:2007vw, Alioli:2010xd, Hamilton:2013fea, Hamilton:2015nsa, STDM-2012-23, Nason:2009ai, Luisoni:2013cuh, Hartanto:2015uka, Jager:2015hka}, Dalitz events (\textit{i.e.} the decay in which one of the two photons is off-shell) are not included.
The decays of bottom and charm hadrons are simulated using the \EVTGEN[1.6.0] programme~\cite{Lange:2001uf}.

In this search, di-photon background processes are of two types: resonant production of a single Higgs boson decaying into $\gamma\gamma$ and continuum non-resonant di-photon production in association with jets.
Background from inclusive production of events with either one photon and one jet ($\gamma j$) or two jets ($jj$), where one or two jets mimic the signature of one or two photons, respectively, is taken into account using a data-driven approach.
All the main single Higgs boson production modes are generated, mostly following the same procedure as in Ref.~\cite{HDBS-2021-10}, including ggF, VBF, production in association with a $W$ or $Z$ boson, with a $t\bar{t}$ or $b\bar{b}$ pair, or with a single top-quark.
Dalitz decays of the Higgs boson are not included here either.
Continuum di-photon production in association with jets from quarks of any flavour ($\gamma\gamma$+jets) is also simulated.
An additional sample of simulated continuum di-photon plus two $b$-jets events ($\gamma\gamma+b\bar{b}$) is generated and used for the study of the background di-photon invariant mass distribution, as described in Section~\ref{sec:modelling}.
For the simulation of continuum di-photon events, \SHERPA[2.2.14]~\cite{Bothmann:2019yzt} is used for both the event generation and the parton showering, and the \NNPDF[3.0nnlo] PDF set is adopted.

All simulated signal and background events are passed through a detailed simulation of the ATLAS detector~\cite{SOFT-2010-01} based on \GEANT~\cite{Agostinelli:2002hh},
except BSM signal samples (i.e. where the value of at least one of the coupling modifiers $\kappa_\lambda$, $\kappa_{2V}$ or $\kappa_V$ is not set to 1) and the non-resonant di-photon samples ($\gamma\gamma$+jets and $\gamma\gamma+b\bar{b}$),
which are interfaced to a fast detector simulation based on a parametric description of the calorimeter showers~\cite{SIMU-2018-04}.
The difference between the results obtained with full simulation and with fast simulation is within the statistical uncertainty, when considering the upper limits on $\mu_{HH}$ and the $\kappa_\lambda$ constraints.
The Higgs boson mass is assumed to be 125 \GeV, a value consistent with the latest measurements~\cite{HIGG-2022-20}, in all simulations.
Additional inelastic $pp$ collisions occurring in the same or nearby bunch crossings, referred to as \textit{pileup}, are also simulated and matched to the distribution observed in data, as described in Refs.~\cite{HDBS-2021-10, HIGG-2022-12}.
%


%
\section{Object and event selections}
\label{sec:selections}
%

The results were obtained using the software framework documented in
Ref.~\cite{easyjet}, developed for collision data analysis within the ATLAS
Collaboration.
The analysis flow begins with the selection of events containing exactly two well-reconstructed photons and at least two $b$-jet candidates~\cite{FTAG-2023-05}.
This pre-selection, briefly summarised in Section~\ref{ssec:preselection}, reduces the background while maintaining good signal acceptance times efficiency.
The selected events are then classified into mutually exclusive categories based on the invariant mass of the $b \bar{b} \gamma\gamma$ system and multivariate discriminants.
The event categorisation, described in Section~\ref{ssec:categories}, is designed to optimise sensitivity to the \hhyybb process and to the coupling modifiers \kl and \kvv.

\subsection{Event pre-selection}
\label{ssec:preselection}
To identify $H \to \gamma\gamma$ decays, events were collected with di-photon triggers with nominal transverse momentum (\pT) thresholds of 35~\GeV\ and 25~\GeV\ for the leading and sub-leading photon, respectively, and they satisfy online photon identification requirements~\cite{TRIG-2022-01,TRIG-2018-05}.
In Run~3, from 2023 onwards, with the upgrade of the first-level trigger system, more granular information from EM calorimeters is used, which results in similar performance compared to Run~2, in an environment with a higher level of pileup~\cite{TRIG-2022-01}.
Of all reconstructed collision vertices, the primary vertex (PV) is selected as the one with the highest sum of squared transverse momenta of contributing tracks $\sum \pT^2$~\cite{ATL-PHYS-PUB-2015-026}.
Selecting the correct PV ensures that all reconstructed objects are associated with the hard-scatter collision and improves the $\gamma\gamma$ mass resolution.

Selected events are required to contain two photon candidates in the acceptance of the finely segmented part of the EM calorimeter ($|\eta|<1.37$ or $1.52<|\eta|<2.37$).
Photons are reconstructed from EM calorimeter signals exceeding noise thresholds using dynamic topological clustering, with additional clusters included for photon conversion products when applicable~\cite{EGAM-2018-01}.
The photon candidates are calibrated~\cite{EGAM-2021-02} and identified by an algorithm that relies on the shape of the EM showers reconstructed in the calorimeter~\cite{EGAM-2021-01}.
A \textit{loose} working point, corresponding to high signal efficiency at the cost of reduced background rejection, is used to match to the photon candidates selected by the trigger.
The final selection of photon candidates is made using a \textit{tight} selection, which provides high background rejection at the cost of reduced signal efficiency.
Photon candidates must also satisfy isolation requirements based on the energy flow in the calorimeter and the total transverse momentum of charged-particle tracks from the PV, measured in cones surrounding the photon direction~\cite{EGAM-2018-01}.
The two leading photons satisfying these selections are required to have an invariant mass \myy between 105 and 160~\GeV\ and transverse momenta above 35\% and 25\% of \myy, respectively.
The \myy \textit{sidebands} are defined as data entries within the mass window of $m_{\gamma\gamma} \in$ [105, 160] \GeV, excluding the signal region from 120 to 130 \GeV.

Jets are reconstructed from particle-flow objects built from noise-suppressed positive-energy topological clusters in the calorimeter and reconstructed tracks using the anti-$k_t$ clustering algorithm with the radius parameter $R=0.4$~\cite{Fastjet,Cacciari:2008gp}.
Jet candidates are required to have $\pT>25$~\GeV\ and $|y|<4.5$.
Jets contained in the fiducial acceptance of the inner detector ($|\eta|<2.4$) and with $\pT<60$~\GeV\ must be identified by a `jet-vertex tagger' as originating from the PV~\cite{PERF-2014-03},
while those outside of this region must be identified by a `forward jet-vertex tagger'~\cite{ATL-PHYS-PUB-2019-026}.
To target $H \to \bb$ decays, events are required to contain at least two {\em $b$-}tagged jets, which are {\em central} jets (those in the acceptance of the inner detector $|\eta|<2.5$) that satisfy the criteria of the GN2 $b$-tagging classification algorithm with an efficiency of 85\% for $b$-jets and a misidentification rate of 0.01 (0.20) for
light-flavour (charm) jets in simulated $t\bar{t}$ events~\cite{FTAG-2023-05,FTAG-2019-02,FTAG-2020-08,FTAG-2018-01}.
These loose requirements are chosen to maximise the signal acceptance and correspond to the optimal analysis sensitivity.
If there are more than two $b$-tagged jets in an event, the jets with the highest $b$-tagging scores (\textit{i.e.} the output discriminant of the GN2 algorithm) are selected to form the $H \to \bb$ candidate.
The $b$-tagging score distribution is discrete, so that each bin corresponds to a given $b$-tagging efficiency range.
If two or more jets are in the same bin of the $b$-tagging score distribution, those with the highest transverse momenta are selected as $H \to \bb$ candidate jets.
Two corrections are applied to the energy of the $b$-tagged jets to
account for possible contributions from muons originating from semileptonic
$b$-hadron decays (Muon-in-jet correction) and undetected energy from neutrinos and out-of-cone
effects (PtReco correction)~\cite{HDBS-2021-10}.
These corrections improve the resolution of the invariant mass of the two $b$-jet candidates from the $H \to \bb$ decay (\mbb) by up to 17\%.

Similarly to Ref.~\cite{HIGG-2020-20}, a kinematic fit (KF) is performed to improve the resolution of \mbb and the invariant mass of the system with two photons and two $b$-jets (\mbbyy), in a two-step process.
In the first step, the KF exploits momentum conservation in the transverse plane to correct jet energies and momenta using the di-photon kinematics.
In the second step, an additional term is introduced to further enhance the resolution with which $m_{b\bar{b}\gamma\gamma}$ is reconstructed by constraining $m_{b\bar{b}}$ to the Higgs boson mass through a quadratic term, $(m_{b\bar{b}}-125\ \GeV)^2$, added to the log-likelihood function being minimised to balance the $HH$ system.
Independent detector-response functions of the jet energies and transverse
momenta are computed using simulated $HH$ events.
The probability density functions of the $x$ and $y$ components of the total transverse momentum of the di-photon system and all jets in the event, as derived from simulation, are used in the KF\@.
The fit output is a correction factor $\alpha$ applied to both the energy and the $\pT$ of the jet. The $\alpha$ factor is evaluated for each jet in the event.
The photon momenta are not corrected by
the KF given the excellent detector resolution and to avoid biasing the $m_{\gamma \gamma}$ distribution. The corrected jet
momenta from the second step are used through the rest of the analysis
while, in the multivariate event discrimination, the \mbb value from the first step of the KF is used to avoid distortions of the background.
After the second KF step, the resolution on \mbbyy is improved by 46\%.
Figure \ref{fig:kinfit} shows the \mbb distribution for data and for simulated signal and background processes.
The $HH$ and $ZH$ simulated events are shown both before and after applying the Muon-in-jet and PtReco corrections, as well as after the first-step KF\@.
The suite of corrections and the KF improve the \mbb resolution by up to 27\%.

\begin{figure}
\centering \includegraphics[width=0.8\textwidth]{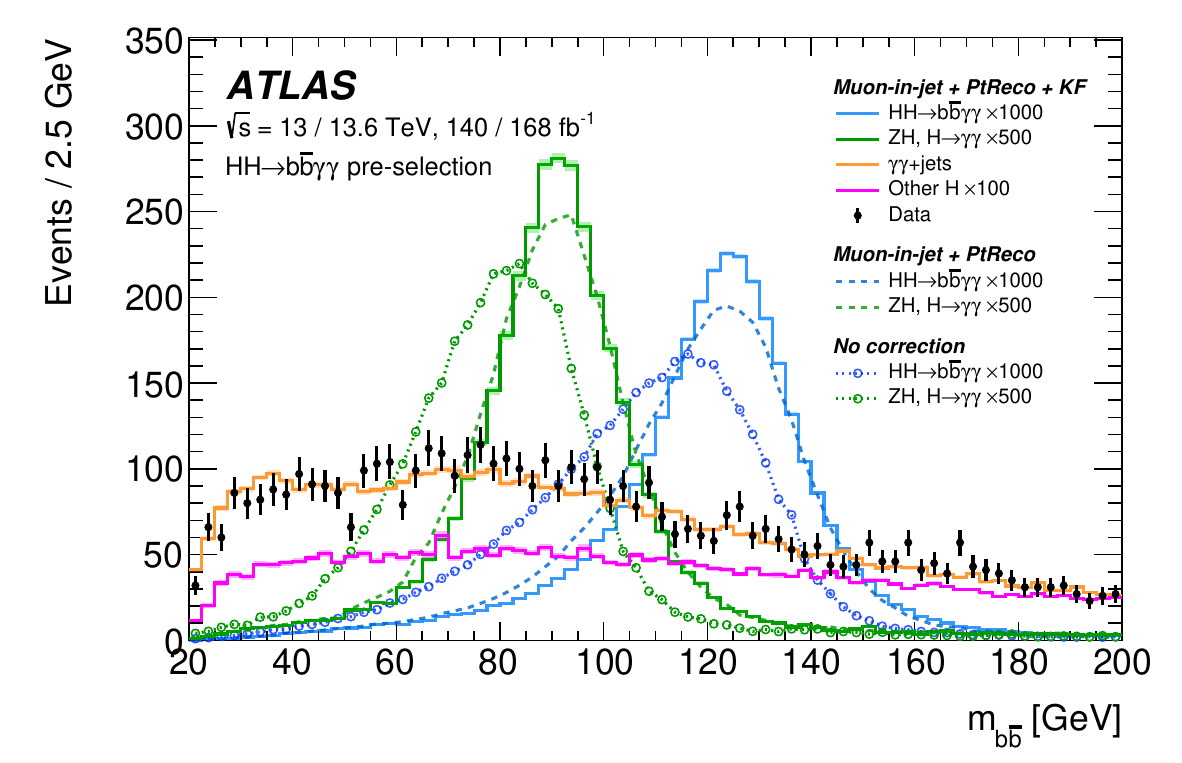}
\caption{ Dijet invariant mass distribution of $b$-tagged jets
(\mbb) after the analysis pre-selection, shown for data as well as
for simulated \hhyybb signal, $ZH$ ($H \to \gamma\gamma$), $\gamma \gamma$+jets,
and other single Higgs boson backgrounds (denoted by `other H').
Also shown are the \hhyybb and  $ZH$ ($H \to \gamma\gamma$) distributions
after applying the Muon-in-jet, PtReco, and the first-step  kinematic fit (KF) corrections.
The shaded regions on the histograms indicate the size of the statistical uncertainties.
All simulated distributions are scaled by the factors shown in the legend, except for $\gamma \gamma$+jets, which is normalised
to the sidebands of the $m_{\gamma\gamma}$ distribution in data.
}
\label{fig:kinfit}
\end{figure}

Events with six or more central jet candidates, or with one or more isolated lepton (electron or muon) candidates with $\pT>10$~\GeV\ and satisfying the lepton \textit{medium} identification criteria~\cite{EGAM-2021-01,MUON-2018-03} are rejected to suppress the background from $t\bar{t}H(\gamma\gamma)$ and inclusive $t\bar{t}$ production.
No requirement is made on the number of \textit{non-central} jets (those beyond the acceptance of the inner detector $|\eta| \geq 2.5$).

The efficiency of the event pre-selection is 17\% and 11\% for SM ggF $HH$ and SM VBF $HH$ events, respectively, with the largest losses arising from the di-photon acceptance and $b$-jet requirements.
The requirement of having at least two $b$-jets is less efficient for the VBF $HH$ events (35\%) than for the ggF $HH$ events (49\%).
A total of $2\,795$ and $3\,321$ events are pre-selected in the Run-2 and Run-3 data, respectively.
According to the SM prediction, a total of about five $HH$ events are expected after the analysis pre-selection.
In this data sample, approximately 74\% of the continuum background events contain two prompt photons, 24\% contain one prompt photon and one misidentified jet, and 2\% contain two misidentified jets, as determined with data-driven techniques~\cite{HIGG-2020-16}.
The fractions of these components in Run~2 and Run~3 are consistent within the statistical uncertainties.
Figure~\ref{fig:myyHist} shows the \myy distributions of data and simulated events after pre-selection for Run~2 and Run~3.

\begin{figure}
\centering
\subfloat[]{ \includegraphics[width=0.52\textwidth]{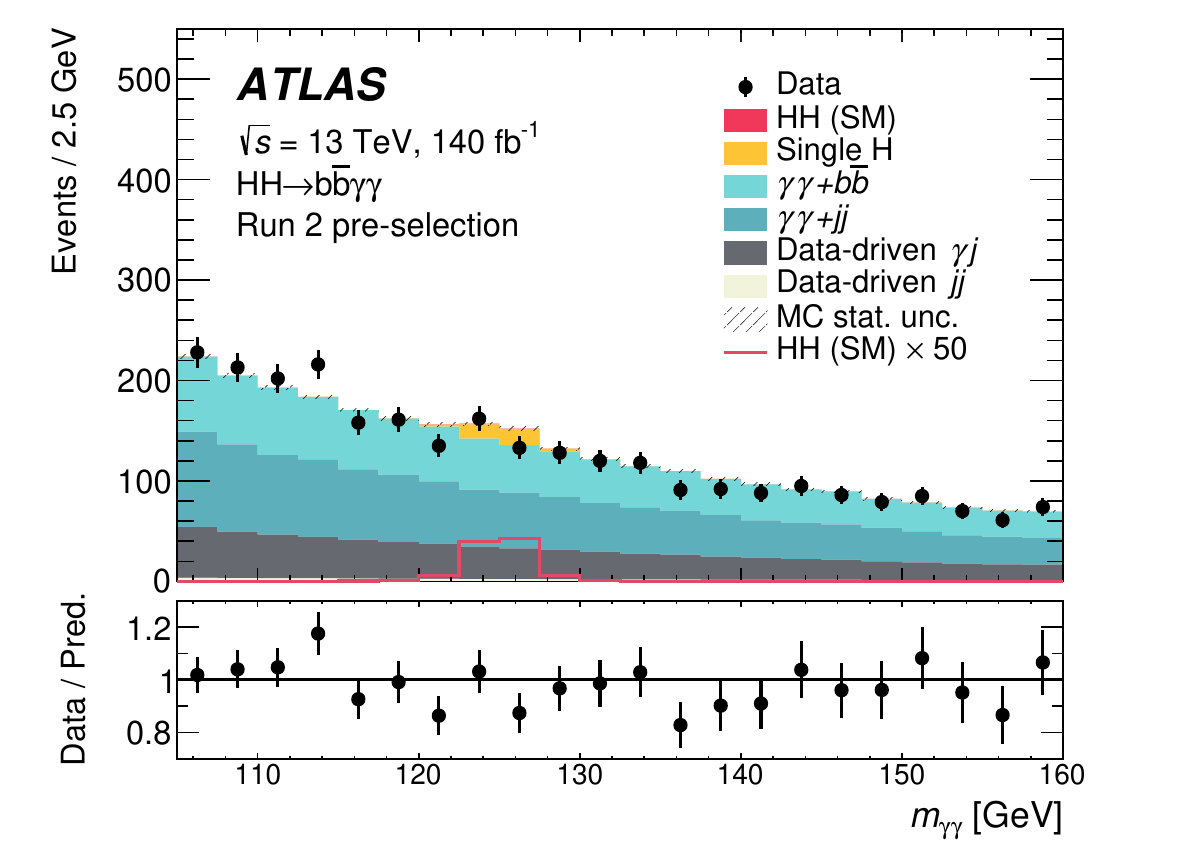}}
\subfloat[]{ \includegraphics[width=0.52\textwidth]{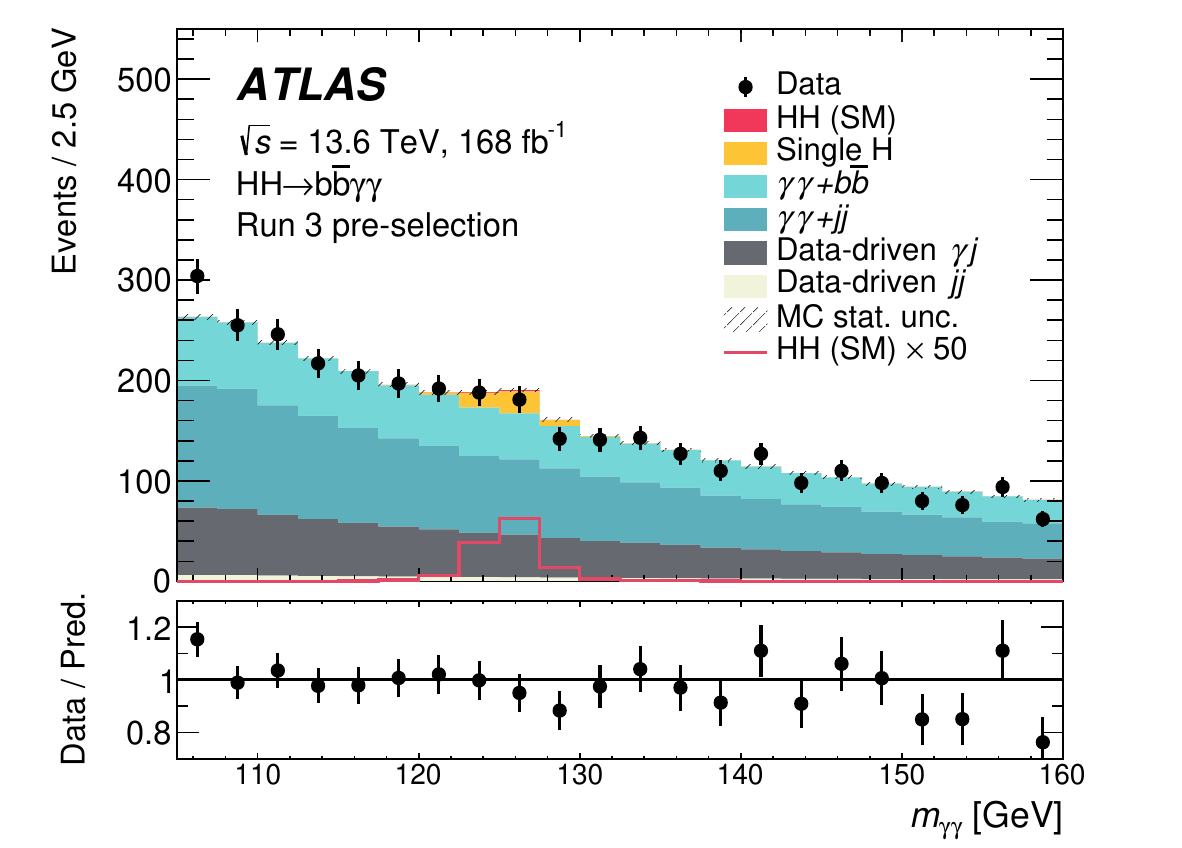}}
\caption{Distributions of the di-photon invariant mass for events in data compared with the sum of the expected signal and background contributions after pre-selection for (a) Run~2 and (b) Run~3. The relative fractions of the simulated $\gamma\gamma+b\bar{b}$ and $\gamma\gamma+jj$ (where $j$ is a light-flavour or gluon jet) events are determined from the generator-level flavour decompositions of the inclusive $\gamma\gamma$+jets sample. The various continuum background components are then normalised using a data-driven photon purity measurement~\cite{HIGG-2020-16} such that their sum is consistent with the data in the sidebands. The $HH$ signal is shown as a filled histogram stacked on top of the backgrounds, normalised to the SM prediction. It is also shown unstacked as an unfilled histogram, scaled by the factor indicated in the legend to enhance visibility.}
\label{fig:myyHist}
\end{figure}

\subsection{Event categories}
\label{ssec:categories}
Anomalous values of the coupling modifiers \kl and \kvv not only affect the $HH$ signal production cross section but also modify the kinematic properties of the $HH$ system.
For example, \kl values different from unity mostly result in a softer $HH$
invariant mass spectrum~\cite{DiMicco:2690841}, while anomalous values of \kvv tend to harden the spectrum.
The events are therefore classified in two regions based on the improved four-body invariant mass $\myybb^{*} = m_{\yybb} - (m_{\bb} - 125~\GeV) - (m_{\gamma\gamma} - 125~\GeV)$:
a \textit{high-mass} region and a \textit{low-mass} region are defined by setting a 350 \GeV boundary on $\myybb^{*}$.
The use of $\myybb^{*}$ over $\myybb$ improves the signal mass resolution by up to 10\% due to the cancellation of detector resolution effects~\cite{HDBS-2018-34}.

In each of the two $\myybb^{*}$ regions, a dedicated boosted-decision-tree (BDT) discriminant is trained to distinguish $HH$ signals from the background arising from $H\to\gamma\gamma$ decays in single Higgs boson production and from the continuum
$\gamma\gamma$+jets events.
The training is performed with the \AtlasMC{XGBoost} programme~\cite{Chen:2016:XST:2939672.2939785} using simulated Run-2 and Run-3 collision events combined according to the SM expectations.
As explained below, Run-2 and Run-3 data events are classified separately.
In the high-mass region, the signal samples used for training include a mixture of SM ggF and VBF $HH$ events, as well
as five BSM samples of VBF $HH$ events with $(\kl, \kvv, \kv)$: (0, 1, 1), (10, 1, 1), (1, 1.5, 1), (1, 3, 1), and ($-5$, 1, 0.5).
The BSM VBF $HH$ samples are used for the training in the high-mass region to increase the sensitivity to $\kvv$.
In the low-mass region, the signal samples consist of BSM ggF $HH$ events corresponding to $\kl = 10$ and $\kl = 5$, plus the five BSM VBF $HH$ samples already used in the high-mass region.
The mixture of BSM samples used for the training in both the high-mass and low-mass regions cover a wide range of anomalous values of $\kl$ and $\kvv$.
Variations of the order of unity on the $\kl$ value used in the training samples have a negligible impact on the final results.
In both regions, the background samples used for the training include single Higgs boson production and continuum di-photon production in association with jets.
The relative weights assigned to each input sample in the mixture of the training dataset, as well as the values of the \AtlasMC{XGBoost} hyperparameters, were tuned in Ref.~\cite{HDBS-2021-10} and their values were found to be optimal for this analysis.

The BDT discriminants use the same input variables as in Ref.~\cite{HDBS-2021-10}.
They include kinematic properties of the two photon
and two $b$-jet candidates, as well as event-level kinematic quantities such as $\mbbyy^*$.
The input variables are chosen in order to avoid distortions of the $\myy$ distribution, which is used as the analysis observable in the final fit to the data, after placing a selection requirement on the BDT output score.
In particular, the $\pT$ of each of the two photons normalised to $\myy$ are included in the input.
Additional variables include,
for events with more than two jets, the $\pT$, $\eta$, $\phi$, and discrete $b$-tagging scores of the third and fourth jets.
Events with at least four jets can arise from VBF $HH$ production, in which the two final-state jets have high transverse momenta and are separated by a large rapidity gap.
VBF jet candidates are selected as those belonging to the pair with the largest invariant mass among all possible pairs in the event excluding the $H \to \bb$ jet candidates.
Their invariant mass and pseudorapidity difference are then used as additional input variables in the BDTs.
In simulated SM VBF $HH$ events with at least four jets, the correct VBF jet pair is identified in 88\% of events.

The two BDT scores are used to define output categories with different signal purities.
They are designed to maximise the overall number-counting significance~\cite{Cowan:2010js} for a benchmark signal
using $HH$ and background yields in the di-photon invariant mass range
$120~\GeV<m_{\gamma\gamma}<130$~\GeV.
In the high-mass region, the signal yield is computed from the sum of the
expected SM ggF and VBF $HH$ contributions, while in the low-mass region,
the benchmark signal is defined using a mixture of two ggF and VBF $HH$ samples with large values of $\kl$ in order to cover a wide range of BSM scenarios.
During this optimisation process, each category is required to contain at least eleven expected continuum background events from the region $105~\GeV<\myy<160$~\GeV
in the combined samples of Run-2 and Run-3 simulated events.
Under the assumption of a Poisson distribution, eleven simulated events in this combined sample correspond to a data sample expected to contain at least two events.
This ensures a sufficient number of events for constraining the shape of the di-photon invariant mass distribution of the continuum background when performing the fit to the data.
This approach allows a more stringent selection requirement on the BDT score, resulting in a 10\% improvement in signal significance compared to a strategy with separated trainings for Run~2 and Run~3, and for which the requirements on continuum background simulated events are applied to the corresponding BDTs individually.

After the optimisation, three categories (labelled `High Mass $i$', $i=1..3$) in
the high-mass region and four categories (labelled `Low Mass $i$', $i=1..4$)
in the low-mass region are defined based on the corresponding
BDT scores, with a higher category index $i$
corresponding to higher BDT scores and more signal-like events.
The addition of further categories did not improve the results significantly.
Events with a BDT score lower than the threshold defining the category
with the lowest index (\textit{i.e.} the category with the lowest signal sensitivity) are discarded.
Data events from the inclusive signal region are thus classified in
fourteen (two sets of seven) exclusive signal regions based on the value of
$\myybb^{*}$, the corresponding BDT scores and the data-taking period (\textit{i.e.} Run~2 or Run~3).

The BDT discriminant distributions in the high- and low-mass regions
observed in data in the \myy sidebands for Run 2 and Run 3 combined are shown in Figure \ref{fig:bdt_score}.
The BDT score distributions
for the dominant non-resonant background from the $\gamma\gamma$ + jets sample,
the resonant single Higgs boson background, and
the combined ggF and VBF $HH$ signals for different values of
$\kl$ and $\kvv$, all taken from simulation, are also shown.
The BDT categories are indicated by vertical dashed lines.
In total, 293 and 342 events are retained in the Run-2 and Run-3 datasets, respectively.
Among these, 239 Run-2 and 269 Run-3 events lie in the \myy sidebands.

\begin{figure}[htbp]
\centering
\subfloat[]{\includegraphics[width=0.48\textwidth]{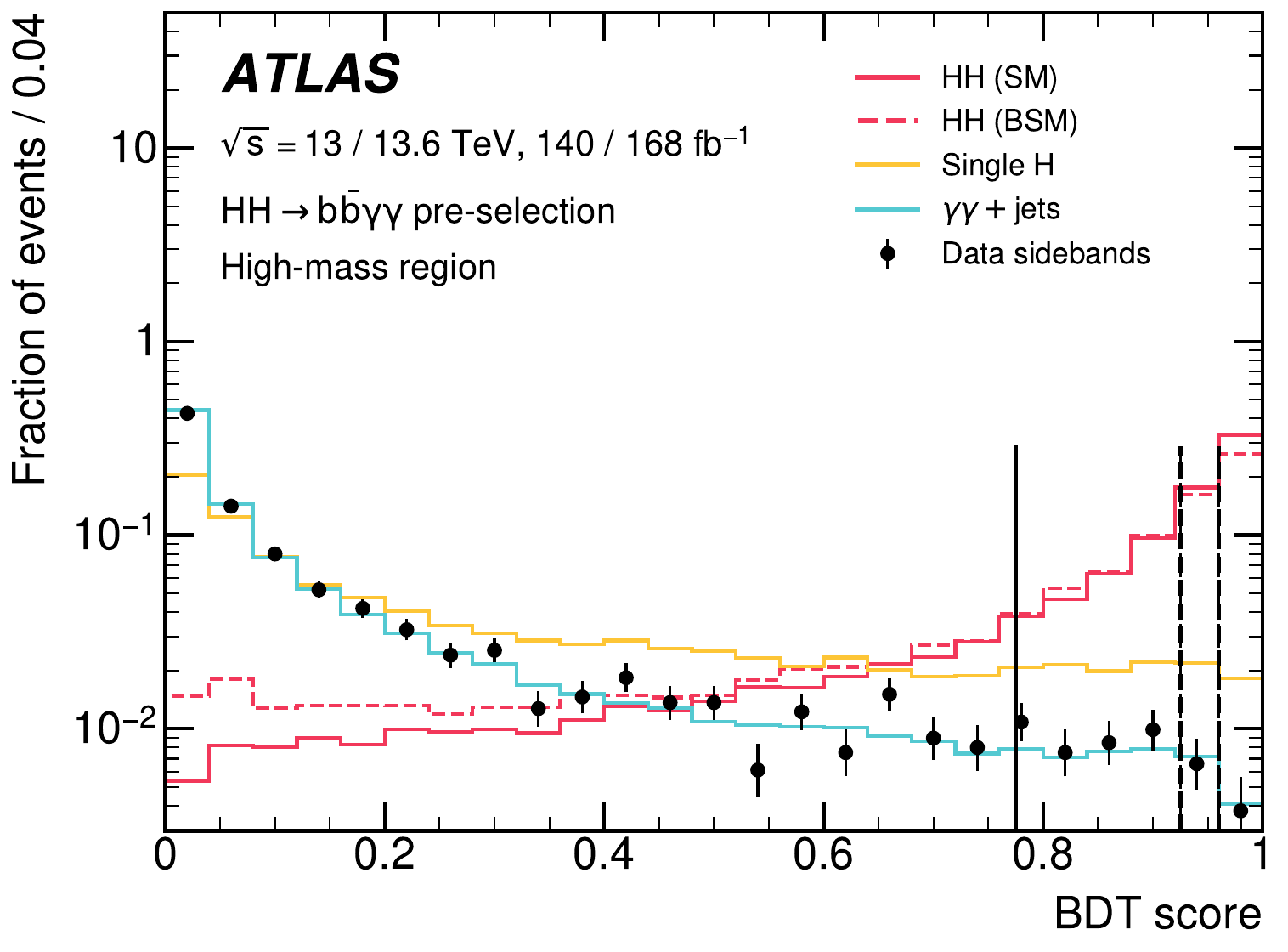}}
\subfloat[]{\includegraphics[width=0.48\textwidth]{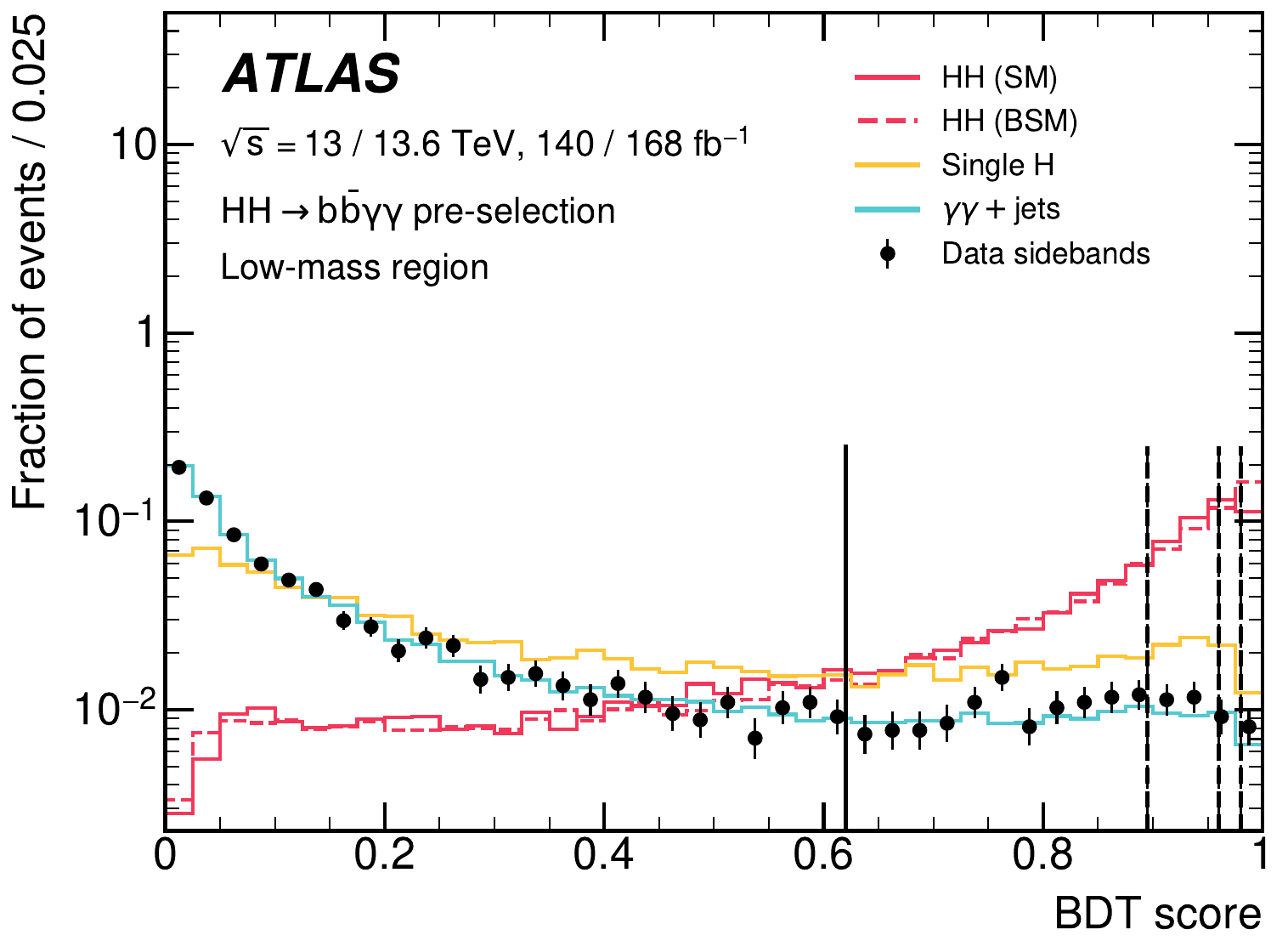}}
\caption{ BDT score distributions using combined Run-2 and Run-3 samples for ggF $HH$ and VBF $HH$ signal events and the two main background components ($\gamma\gamma$ + jets continuum background and single Higgs boson processes) for the (a) high-mass and (b) low-mass regions. Data sideband events are also shown. %
Simulated events are contained in the interval $m_{\gamma\gamma} \in$ [105, 160] \GeV.
All distributions are normalised to unity. The vertical dashed lines
correspond to the thresholds used to define the event categories. Events with a BDT score below the lowest threshold (solid line) are discarded.}
\label{fig:bdt_score}
\end{figure}

%

%
%
%

%
%

%
%
%
%
%
%
%
%
%
%
%

%
%
%
%
%
%
%
%
%
%
%
%
%
%
%
%
%
%
%
%
%
%
%
%
%
%
%
%
%
%
%
%

%
%
%
%
%
%
%
%
%


%
\section{Signal and background modelling}
\label{sec:modelling}
%

For the extraction of the statistical results, a simultaneous fit is performed to all data in the fourteen event categories.
The signal and non-resonant background yields in each event category
are determined from the combined unbinned fit to the di-photon
invariant mass distributions, while the resonant background yields are fixed to the SM expectation.
The signal yields for ggF and VBF $HH$ production are parameterised as a function of the relevant coupling modifiers by performing a linear combination of the relevant simulation samples.
The differential VBF $HH$ cross section can be written as a polynomial of the coupling modifiers, whose coefficients are determined using six reference samples for $(\kl, \kvv, \kv)$:
(1, 1, 1), (1, 1, 0.5), (10, 1, 1), (1, 1.5, 1), (2, 1, 1), and ($-5$, 1, 0.5).
Similarly, the ggF $HH$ cross section is expressed as a function of \kl using three samples with $\kappa_\lambda=0$, $1$, $5$.

The signal and background $\myy$ distributions in each category
are independently modelled with analytical functions.
For signal and resonant background events, double-sided Crystal Ball functions~\cite{dscb1,HIGG-2013-08,HIGG-2014-04} are used.
The nominal shape parameters are obtained from fits to simulated SM $HH$
events, with ggF and VBF modes weighted according to their SM cross sections and summed.
In the final fit to data, the Gaussian core parameters, describing the peak and width of the distribution, are allowed to float within the uncertainties arising from the photon energy calibration,
while the tail parameters are fixed to their nominal values.

Signal shapes are parameterised for Run~2 and Run~3 separately.
The same model is found to describe well single Higgs boson and
Higgs boson pair simulated events for both SM and non-SM coupling modifier values.

The $\myy$ distributions of the non-resonant di-photon background are modelled with exponential functions, whose normalisation and shape parameters are obtained from the fit to the data.
The chosen exponential function in each category is found to describe well the data in the $\myy$ sidebands, as well as the background-only template obtained with
a sample of simulated $\gamma\gamma+b\bar{b}$ events normalised to the data in the $\myy$ sidebands.
The background shape parameters are correlated for each Run-2 and Run-3 category pair (\textit{e.g.} High Mass 1 for Run~2 and High Mass 1 for Run~3) in the fit to the data,
ensuring consistency with the requirement on the minimum expected number of continuum background events to define category boundaries, as described in Section~\ref{sec:selections}.
Tests using the large sample of simulated $\gamma\gamma+b\bar{b}$ continuum background events, as well as from a comparison of data-sideband distributions in the low-mass and high-mass regions, indicate good compatibility between the Run-2 and Run-3 shape parameters at the current level of precision of this analysis.
Furthermore, no significant bias on the fitted signal is observed even when enhancing the difference between Run-2 and Run-3 background shapes by a factor of ten compared to those seen in simulation.
Finally, the impact of correlating or decorrelating the parameters of the analytical model describing the shape of the background is found to be negligible.
Normalisation parameters of the background functions are not correlated.


%
\section{Systematic uncertainties}
\label{sec:systematics}
%

Systematic uncertainties affect both the di-photon invariant mass distribution of $HH$ signal and single Higgs boson background. However, their impact remains relatively minor compared with the statistical uncertainty, which dominates due to the limited number of data events and the use of data-driven methods for modelling the dominant continuum background.

The \textit{spurious signal} is defined as the maximum absolute value of the bias on the fitted signal yield.
To estimate it, multiple signal-plus-background fits are performed on a background-only template.
These fits are repeated using different background models, such as exponential and power-law functions.
In addition, different assumptions are made for the signal mass, with $m_H$ varied between $123$ and $127~\GeV$ in steps of $0.5~\GeV$.
The background-only template is constructed without accounting for misidentified photon backgrounds, as their contribution is found to have a negligible impact on the spurious signal estimate.
For each of the background functions tested, the spurious signal is smaller than 20\% of the statistical uncertainty in the expected fitted signal yield.
The spurious signal value found in each category is taken as a systematic uncertainty in the signal yield.

The uncertainty in the full Run-2 integrated luminosity amounts to 0.83\%~\cite{DAPR-2021-01}. The uncertainty in the Run-3 luminosity is estimated to be 4\% and is uncorrelated with that of \RunTwo.

The uncertainty in the efficiency of the di-photon trigger in \RunTwo and the 2022 dataset is estimated by using radiative $Z$ boson decays and events from the pre-scaled lower threshold trigger~\cite{TRIG-2018-05}, while for 2023 and 2024 the same trigger correction factors as in 2022 are employed, based on a study of the trigger efficiencies in data and simulation over the different years of the Run-3 data-taking.
In \RunTwo, the photon identification and isolation uncertainties are determined using control samples of prompt photons from photon+jet production, radiative $Z \rightarrow \ell\ell\gamma$ decays, and electrons from $Z \rightarrow ee$ decays. In the latter case, electrons are used as a proxy for photons due to their similar calorimeter signatures~\cite{EGAM-2018-01}.
The uncertainties in the photon energy scale and resolution in \RunTwo are determined from control samples of electrons from $Z$ boson decays and of photons from radiative $Z$ boson decays
\cite{EGAM-2018-01}.
The corresponding uncertainties for \RunThr, in particular for the photon energy scale and resolution, are extrapolated from \RunTwo~\cite{HIGG-2022-12}.

The photon energy scale and resolution uncertainties affecting the peak and
the width, respectively, of the signal and single Higgs boson background
are estimated with both the SM ggF and VBF $HH$ samples. The peak position
has an uncertainty of less than 0.7\% in all categories, while the uncertainty in the width is less than 21\%.
The photon energy scale and resolution effects on the yields are controlled by the same nuisance parameters as those on the \myy shape.
The systematic uncertainties affecting the yields of processes with at least one Higgs boson are estimated separately for the ggF and VBF $HH$ signals and the single Higgs boson production.
For ggF and VBF $HH$ production, experimental systematic uncertainties are estimated with the SM samples and applied to all $HH$ configurations.

The uncertainties in the jet energy scale and resolution are estimated by using control samples in which jets recoil against well-calibrated reference objects, such as a photon or a $Z$ boson, or other already calibrated jets~\cite{JETM-2018-05}.
An additional uncertainty is introduced in the simulation to account for variations in response among \bjets, light-quark jets, and gluon jets.
The flavour tagging uncertainties include those arising from the efficiencies and misidentification rates. They are estimated in \ttbar events for $b$- and $c$-jets, and in $Z$+jets events for light-quark and gluon jets~\cite{FTAG-2019-07,FTAG-2020-08,FTAG-2019-02}.

For the ggF and VBF $HH$ signal yield parameterisation, non-closure uncertainties are derived from the differences between the yields obtained from the parameterisation and simulated samples with different \kl, \kvv, \kv values. The maximal deviation over all samples is taken as the final uncertainty.
In particular, the high-mass categories have a non-closure uncertainty of $\sim$10\% that is comparable to Ref.~\cite{HDBS-2021-10}.
The numbers in the low-mass regions are higher especially for the VBF $HH$ sample, reaching up to $\sim$30\%. However this does not noticeably affect the sensitivity, as VBF is subdominant to ggF in \kl measurements.

Theory uncertainties in the analysis impact both the acceptance and the cross sections. A key component of these is a 100\% uncertainty applied to the yields of single Higgs boson production modes (ggF, VBF, and $WH$), motivated by studies of heavy-flavour production in association with top-quark pairs~\cite{TOPQ-2017-12,TOPQ-2012-16}, $W$ boson production in association with \bjets~\cite{STDM-2012-11}, and confirmed by measurements of single Higgs boson production in association with \bjets~\cite{HIGG-2018-29,HIGG-2019-13}.
Other theory uncertainties originate from missing higher-order terms in perturbative QCD calculations and from the limited knowledge of PDFs and of the value of the strong coupling constant \alphas.
The uncertainty due to missing higher-order terms is evaluated by independently varying the renormalisation and factorisation scales.
In the fit, all the theory uncertainties are correlated between \RunTwo and \RunThr, but uncorrelated between processes.
These scale uncertainties are calculated separately for \RunTwo and \RunThr, in each category and for each process: for VBF production, they amount to $0.04\%$, while for ggF production, with the addition of the top-quark mass scheme uncertainty included according to Ref.~\cite{Baglio:2020wgt}, they amount to $+6\%$ and $-23\%$.

The PDF and $\alpha_s$  uncertainties are estimated separately for \RunTwo and \RunThr and separately for different processes (single Higgs boson, ggF $HH$ and VBF $HH$ events).
The PDF+\alphas uncertainties for the ggF (VBF) $HH$ cross sections amount to $3.0\%$ ($2.1\%$).
These uncertainties are evaluated for the SM case and their magnitude is parameterised in terms of \kl.
The parton shower model uncertainty is estimated
by using alternative simulation samples. In particular, for the $HH$ and single Higgs boson processes the alternative samples use Herwig 7~\cite{Bahr:2008pv,Bellm:2015jjp} for parton showering.

\begin{table}[ht!]
\caption{Impact of the various systematic uncertainties in the observed $\mu_{HH}$ measurement. For each uncertainty group, the error on $\mu_{HH}$ is obtained from a fit with only those corresponding nuisance parameters floating and all others fixed to their best-fit values.
The impact is then calculated from the subtraction in quadrature of this error and the error when all nuisance parameters are fixed.
The up (down) columns indicate the impact on the upper (lower) $\mu_{HH}$ error.
When the up and down impacts are found to be compatible within 30\%, the values are symmetrised in the table for better readability.
Asymmetric impacts primarily result from the asymmetric QCD scale + $m_{\text{top}}$ signal uncertainty, and the fact that the lower endpoint of the 68\% CL error on the signal strength is close to 0.
The heavy-flavour content uncertainty in the table refers to single Higgs boson production. }
\label{tab:sys}
\centering
\begin{tabular}{lC{1.5cm}C{1.5cm}}
\toprule
\multirow{2}{*}{\textbf{Source of systematic uncertainty}} & \multicolumn{2}{C{4.5cm}}{\textbf{\hspace{0.2cm}Relative impact [\%]}} \\
& \multicolumn{2}{C{4cm}}{\textbf{ Up \hspace{0.8cm} Down}} \\
\toprule
\textbf{Experimental} \\
\ \ Photon energy scale        & \hspace{0.69cm}$+22$ & $-34$ \\
\ \ Photon energy resolution   & \hspace{0.69cm}$+14$ & $-7.5$ \\
\ \ Photon efficiency          & \hspace{0.69cm}$+13$ & $-2.5$ \\
\ \ Jet                        & \hspace{0.69cm}$+10 $ & $-6.9$ \\
\ \ Luminosity                 & \hspace{0.75cm}$+6.7$ & $-1.2$ \\
\midrule
\textbf{Theory} \\
\ \ QCD scale + $m_{\text{top}}$, PDF+$\alpha_S$	       & \hspace{0.69cm}$+35$ & $-5.2$ \\
\ \ $\mathcal{B}$($H\rightarrow \gamma\gamma, b\bar{b}$)&  \hspace{0.69cm}$+10 $ & $-2.1$ \\
\ \ Parton showering model                              & \multicolumn{2}{C{3.8cm}}{$\pm 15$} \\
\ \ Heavy-flavour content                               & \multicolumn{2}{C{3.8cm}}{$\pm 30$} \\
\midrule
\textbf{Background model} \\
\ \ Spurious signal & \multicolumn{2}{C{3.9cm}}{$\pm 7.2$} \\
\bottomrule
\end{tabular}
\end{table}

In addition, branching ratio uncertainties are taken into
account~\cite{deFlorian:2016spz}. Specifically, for all $HH$ and single
Higgs boson processes, a 2.9\% uncertainty is applied to the $H \rightarrow
\gamma\gamma$ branching ratio. An additional 1.7\% uncertainty from the $H
\rightarrow b \bar{b}$ branching ratio is included for the $HH$ signal.

The impact of the various systematic uncertainties on the observed $\mu_{HH}$ measurement is shown in Table~\ref{tab:sys}.
For each uncertainty group, the error on $\mu_{HH}$ is obtained from a fit with only those corresponding nuisance parameters floating and all others fixed to their best-fit values.
The impact is then calculated from the subtraction in quadrature of this error and the error when all nuisance parameters are fixed.
The up (down) columns indicate the impact of the upper (lower) $\mu_{HH}$ error.
When the up and down variations are found to be compatible within 30\%, the impacts are symmetrised in the table for better readability.
Asymmetric impacts primarily result from the asymmetric QCD scale + $m_{\text{top}}$ signal uncertainty, and the fact that the lower endpoint of the 68\% CL error on the signal strength is close to 0.
The heavy-flavour content uncertainty in the table refers to single Higgs boson production.

%


%
%

%
\section{Results}
\label{sec:results}
%

The search results are derived with a simultaneous profile-likelihood fit
to the di-photon invariant mass distribution in all event categories, with a procedure similar to that described in Ref.~\cite{HDBS-2021-10}. The parameters of interest are fitted together with a set
of nuisance parameters representing the systematic uncertainties, and
the normalisation and shape of the continuum background. Uncertainties and confidence
levels are derived using the asymptotic approximations~\cite{Cowan:2010js},
which agree within 10\% with those from pseudo-experiments.
Upper limits on the signal strength are derived using the CL$_\text{s}$
approach~\cite{Read:2002hq_custom}.
It is always assumed
that the Higgs boson satisfies exactly all SM predictions
apart from the parameter under study.

\begin{figure}[t]
\centering  \includegraphics[width=0.6\textwidth]{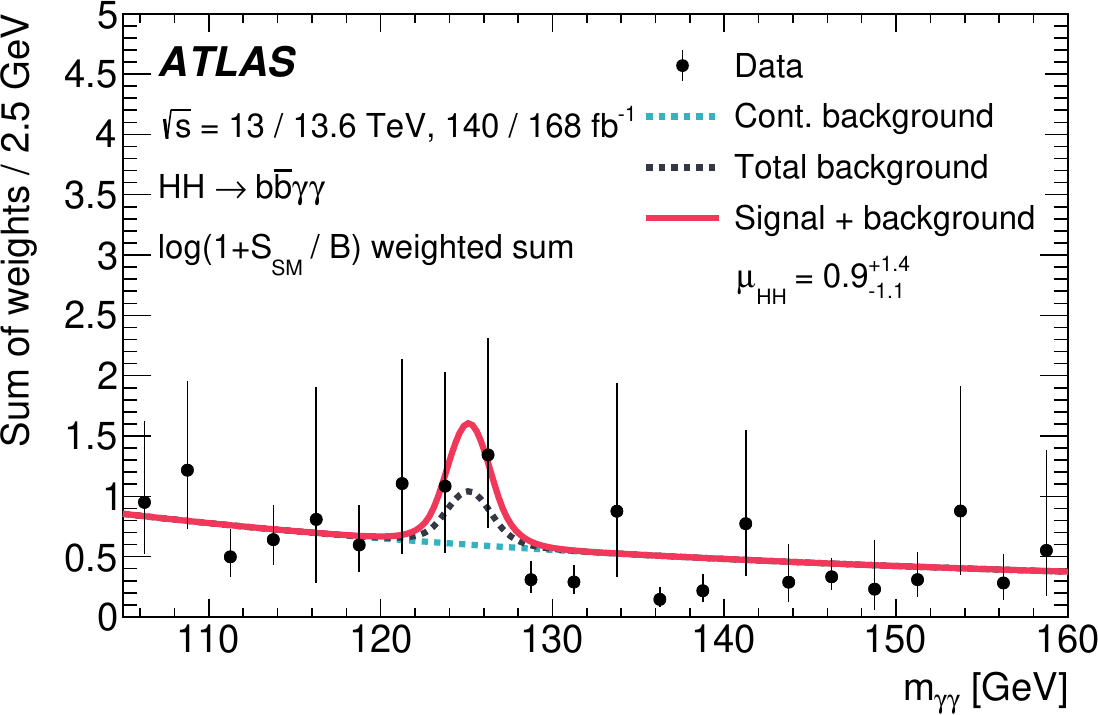}
\caption{Weighted di-photon invariant mass distribution summed over all
categories and the two data-taking periods.  The events in each
category are weighted by $\log(1+S_{\mathrm{SM}}/B)$.
$S_{\mathrm{SM}}$ is the expected signal yield assuming $\mu_{HH}=1$,
while $B$ is the continuum background yield obtained from a fit to the
sidebands plus the single Higgs boson background obtained from
simulation, all in a $\pm 5$\,\GeV window around the Higgs boson mass. The
lines show the fit results for the continuum background only (light
dotted), adding single Higgs boson backgrounds (black dotted) and the
full fit (solid).}
\label{fig:myy}
\end{figure}

Figure \ref{fig:myy} shows the di-photon invariant mass distribution summed
over all categories and the two data-taking periods, where the events in
each category are weighted by $\log(1+S_{\mathrm{SM}}/B)$, where
$S_{\mathrm{SM}}$ is the expected signal yield assuming $\mu_{HH}=1$, while
$B$ is the continuum background yield obtained from a fit to the sidebands
plus the single Higgs boson background obtained from simulation, all in a
5\,\GeV window around the Higgs boson mass.

The best-fit value for the signal strength modifier is
$\mu_{HH} = 0.9^{+1.4}_{-1.1} =
0.9^{+1.3}_{-1.0} \,\mathrm{(stat.)}\,\pm 0.5 \,\mathrm{(syst.)}$
for an expected value of
$\mu_{HH} = 1^{+1.3}_{-1.0}$ in the SM\@.
Fitting the data from Run 2 and Run 3
separately results in $\mu_{HH}$ values compatible with each other with a
$p$-value of 0.40.
The statistical
significance for Higgs boson pair production is $0.78\, \sigma$, where $1.01\,
\sigma$ is expected.
The systematic uncertainty is dominated by the uncertainties in the
heavy-flavour content in single Higgs boson production, the photon energy scale and
resolution, and the calculation of the Higgs boson pair production cross
section.  An upper limit at 95\% CL of $\mu_{HH} < 3.7$ is
derived assuming SM $HH$ kinematics. Assuming no $HH$ production the
expected limit is $\mu_{HH} < 2.6$, while assuming $\mu_{HH} =1$, the
expected limit becomes $\mu_{HH} < 3.7$. Figure \ref{fig:r23mucomp}
compares the separate fit results for Run 2 and
Run 3 as well as their combination in terms of the
significance and the 95\% CL upper limit.
The same definition of the analysis categories is used for all fits.
Compared to Ref.~\cite{HDBS-2021-10} the
significance in $\mu_{HH}$ improved by approximately 20\% due to the new
$b$-tagging algorithm and an additional 5\% comes from the kinematic fit.
The improvement from the additional data is approximately 60\%, with 50\% attributed to increased statistics and 10\% to the optimisation of the analysis categories, made possible by treating the background shapes in the two datasets as correlated.

\begin{figure}[bt]
\centering
\subfloat[]{  \includegraphics[width=0.44\textwidth]{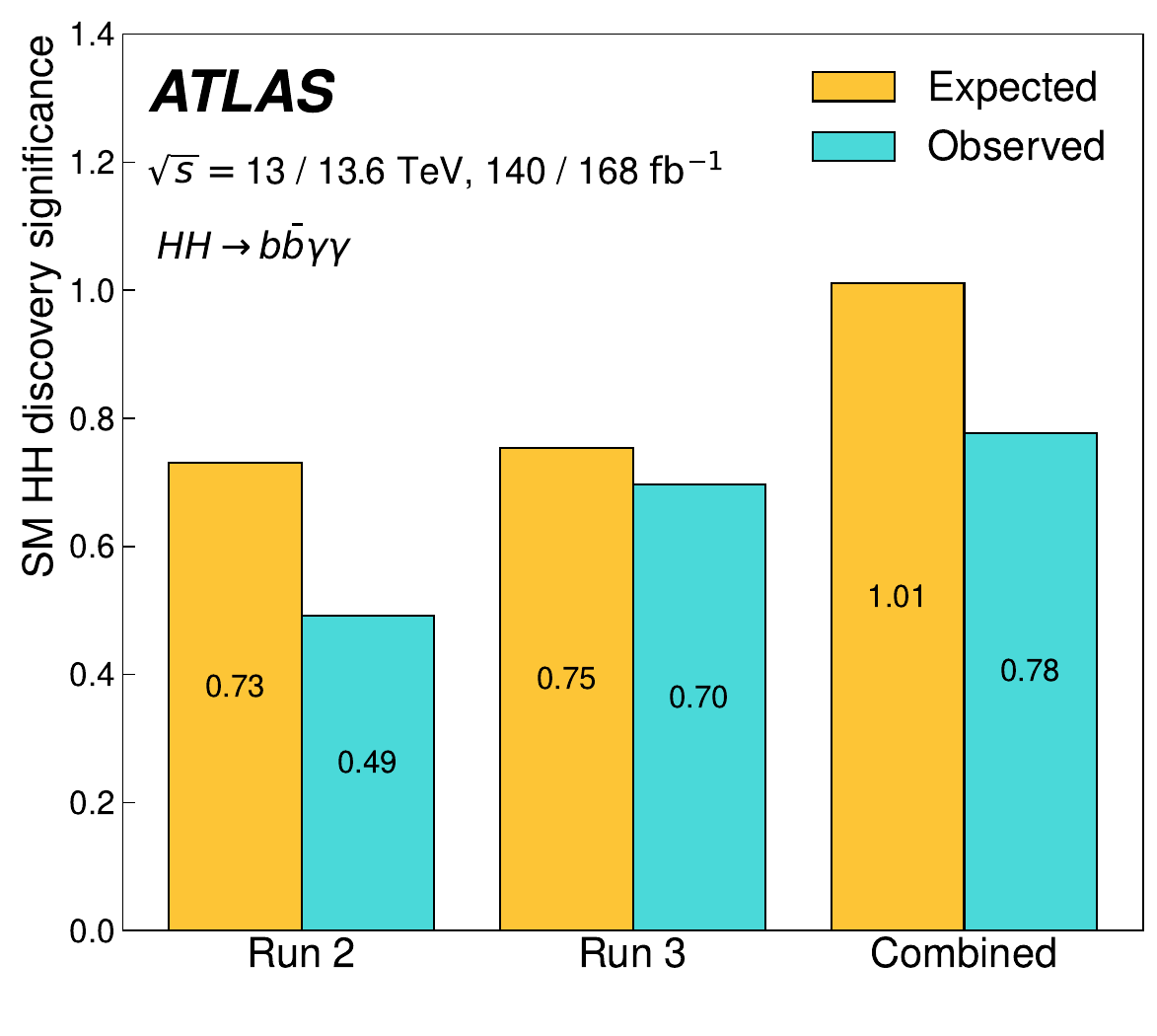}
}
\subfloat[]{
\includegraphics[width=0.52\textwidth]{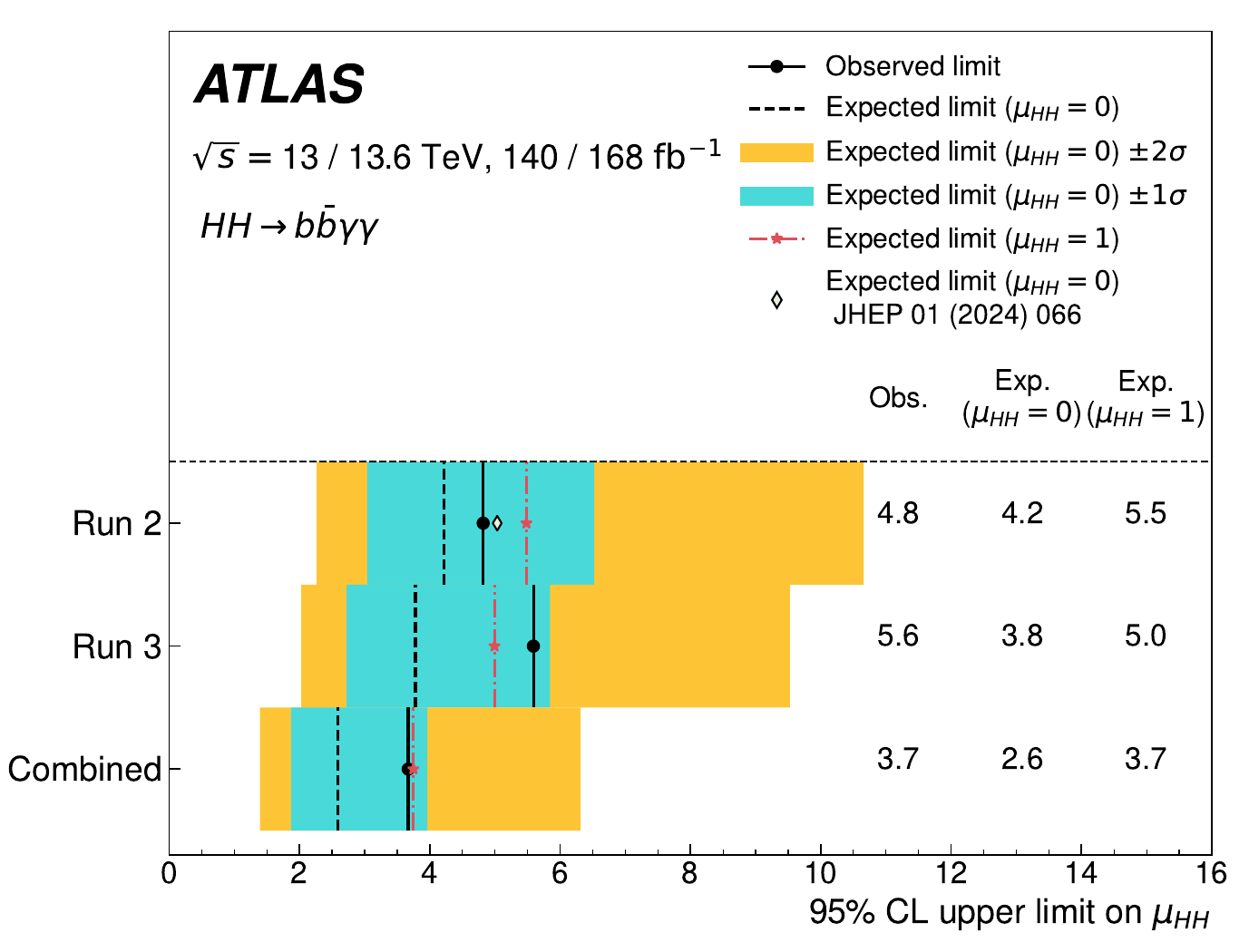}
}
\caption{(a) Observed and expected statistical significance of the $HH$
signal, and (b) the 95\% CL upper limits on the signal strength,
obtained with separate fits to Run-2 and Run-3 data as well as their
combination. When computing the significance or upper limit for one
data-taking period only, $\mu_{HH}$ of the other period is left free
to vary. All other parameters of interest are fixed to their SM
expectation. The expected limit assuming
$\mu_{HH}=0$ from Ref.~\cite{HDBS-2021-10} is also shown in (b). }
\label{fig:r23mucomp}
\end{figure}

Profile-likelihood fits are performed separately for the two coupling
modifiers $\kappa_\lambda$ and  $\kappa_{2V}$, with all others fixed to their SM
values. The likelihood scans are shown in Figures \ref{fig:kappalfit} and
\ref{fig:kappavfit}.
The best fit values are
$\kappa_\lambda = 1.4^{+3.7}_{-1.8}$ and $\kappa_{2V} = 0.9^{+1.2}_{-0.8}$,
compatible with the expected values of
$\kappa_\lambda = 1^{+4.4}_{-1.6}$ and $\kappa_{2V} = 1^{+1.1}_{-0.9}$.
The observed (expected) 95\% confidence intervals are $-1.6 < \kappa_\lambda < 6.6$
($-1.8 < \kappa_\lambda < 6.9$) and $-0.5< \kappa_{2V} < 2.6$
($-0.4 < \kappa_{2V} < 2.6$).  For these coupling modifiers, fits performed
on the Run-2 and Run-3 data separately are compatible with $p$-values of
0.17 for $\kappa_\lambda$ and 0.45 for $\kappa_{2V}$. Compared to
Ref.~\cite{HDBS-2021-10}, the expected width of the 95\% confidence interval
decreases by approximately 20\% for $\kappa_\lambda$ and 30\% for $\kappa_{2V}$, respectively.
The observed confidence intervals are similar
to those in Ref.~\cite{HDBS-2021-10} due to differences in the observed
best-fit values, which in Ref.~\cite{HDBS-2021-10} were driven by a small
deficit of events in the signal region.  A two-dimensional fit with
$\kappa_\lambda$ and $\kappa_{2V}$ as free parameters yields observed and
expected results of
$\kappa_\lambda = 2.8^{+2.2}_{-3.0}$, $\kappa_{2V} = 1.6^{+0.6}_{-1.8}$,
and
$\kappa_\lambda = 1^{+4.7}_{-1.7}$, $\kappa_{2V} = 1^{+1.1}_{-1.3}$,
respectively, where the other parameter is unconstrained when calculating the error.
The central values and
the 68\% and 95\% CL contours are shown in Figure \ref{fig:2dfit}.

\begin{figure}[htb]
\centering
\subfloat[]{
\label{fig:kappalfit}   \includegraphics[width=0.48\textwidth]{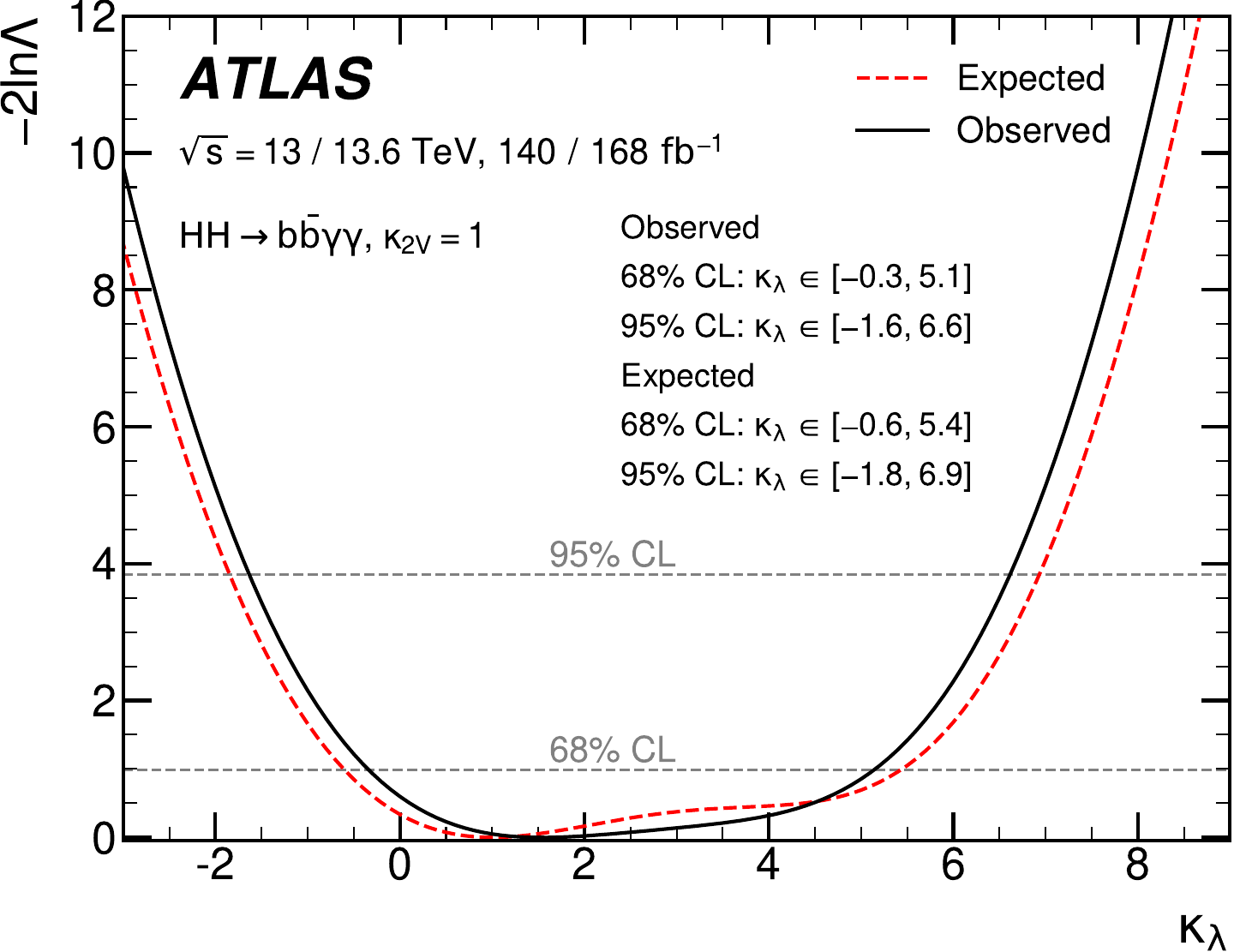}
}
\subfloat[]{
\label{fig:kappavfit}    \includegraphics[width=0.48\textwidth]{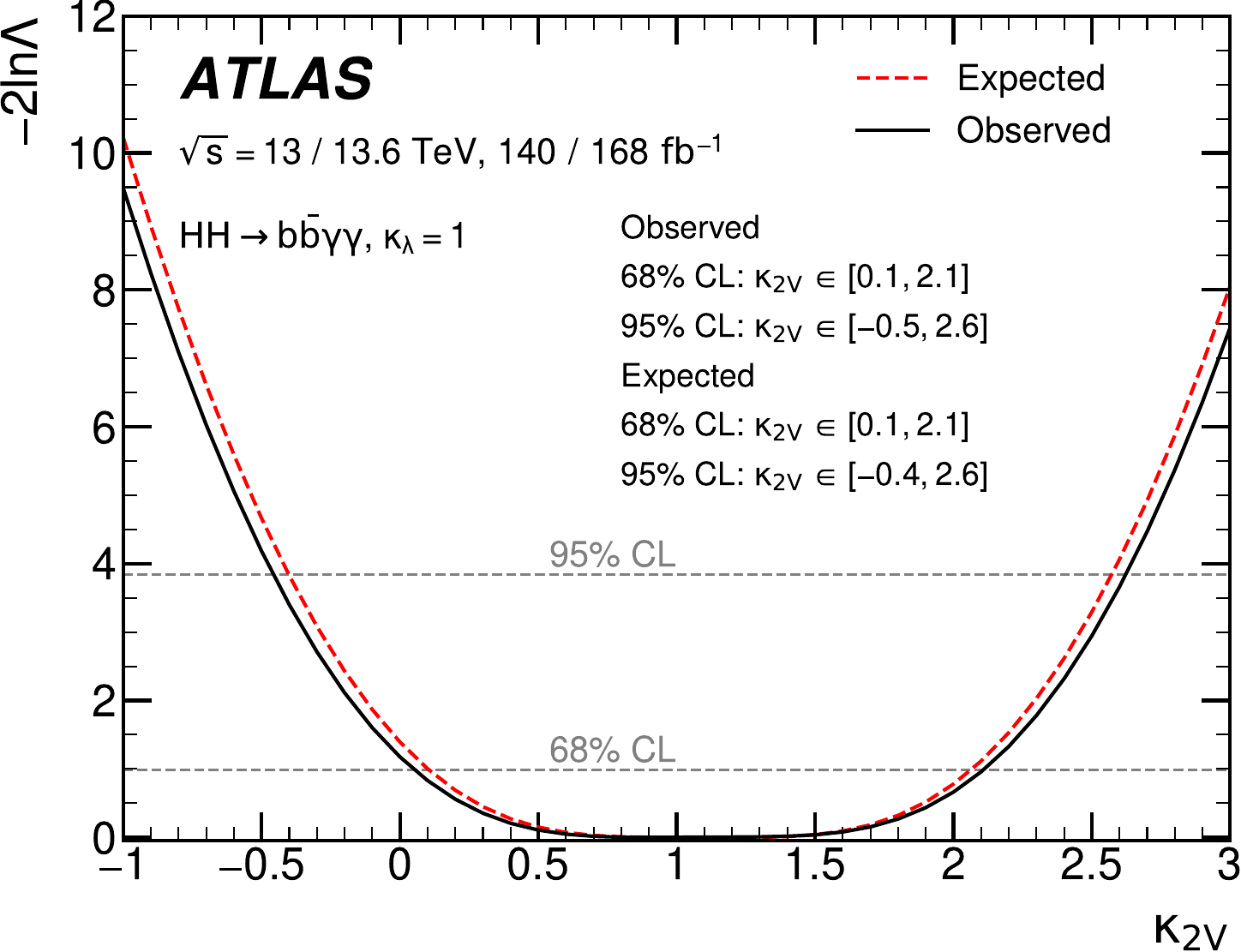}}\\
\subfloat[]{
\label{fig:2dfit} \includegraphics[width=0.48\textwidth]{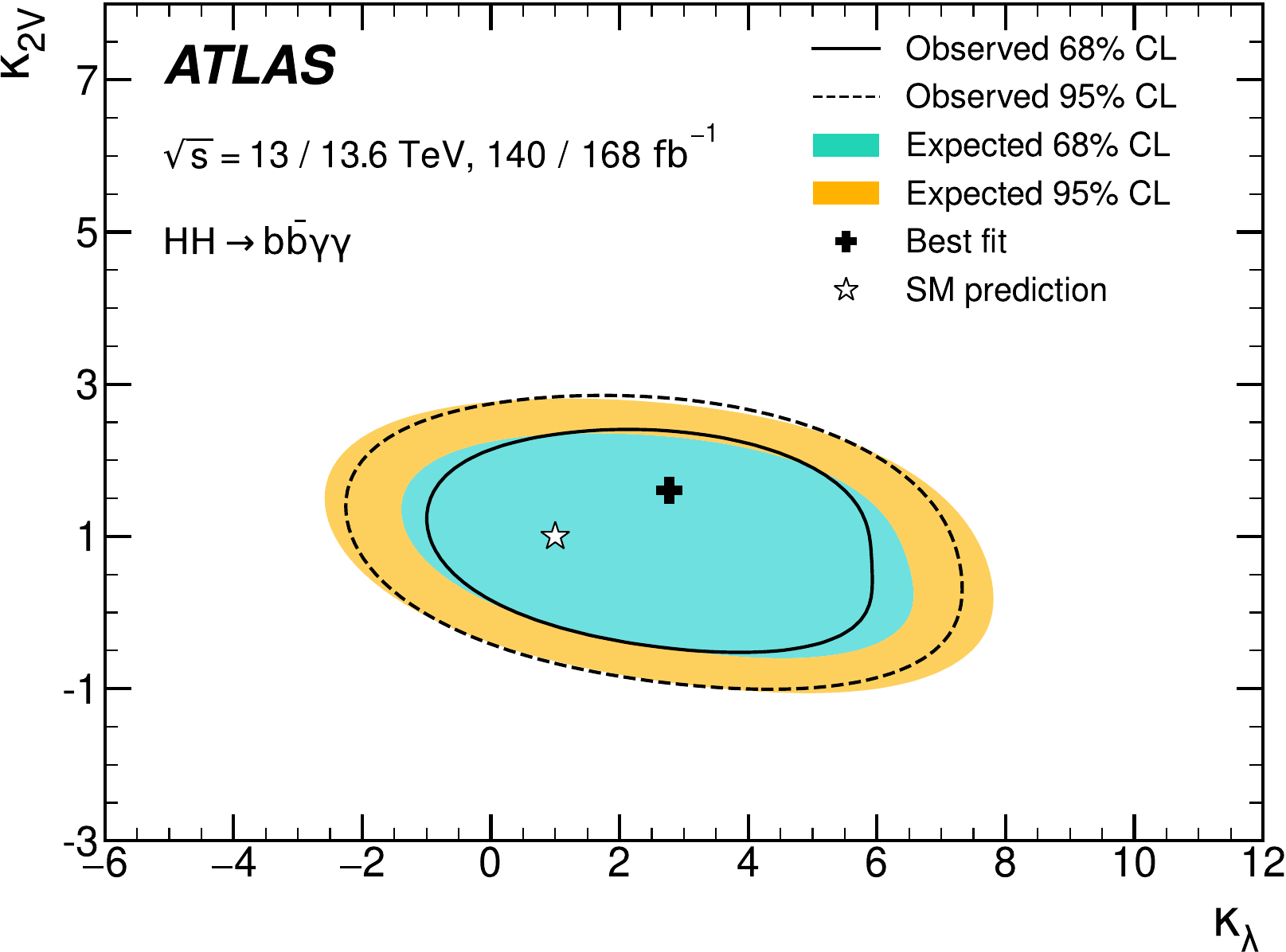}
}

\caption{Observed and expected profile likelihood scan of (a)
$\kappa_\lambda$ and (b) $\kappa_{2V}$. (c) Contours at 68\% CL (solid
line) and 95\% CL (dashed line) in the $(\kappa_\lambda, \kappa_{2V})$
parameter space. The corresponding expected contours are shown by the
inner and outer shaded regions. The SM prediction is indicated by the
star, while the best-fit value is denoted by the cross.  The scans are
performed by varying only the coupling modifier of interest, while all
other relevant coupling modifiers are fixed to unity. For the
two-dimensional fit the individual results are
$\kappa_\lambda = 2.8^{+2.2}_{-3.0}$ and
$\kappa_{2V} = 1.6^{+0.6}_{-1.8}$, where the errors are one dimensional
standard deviations.}
\end{figure}

\FloatBarrier


%
%
%

%
\section{Conclusion}
\label{sec:conclusion}
%

A search for Higgs boson pair production in the $b \bar{b} \gamma
\gamma$ final state is performed.
The proton--proton collision dataset used an integrated luminosity of
308 fb$^{-1}$, and consists of two samples, 140 fb$^{-1}$ at a centre-of-mass energy of
13\,\TeV and 168 fb$^{-1}$ at 13.6\,\TeV, recorded between 2015 and 2024 by the
ATLAS detector at the CERN Large Hadron Collider.
A simultaneous fit to all data in signal regions, categorised according to  the $b\bar{b}\gamma\gamma$ invariant mass, signal purity, and data-taking period, results in a signal strength of $\mu_{HH} = 0.9^{+1.4}_{-1.1}$, compatible with the SM prediction of unity, and corresponding to an upper limit of  $\mu_{HH} < 3.7$. Additional fits constrain the trilinear Higgs and $VVHH$ quartic couplings to
$-1.6 < \kappa_\lambda < 6.6$ and $-0.5< \kappa_{2V} < 2.6$, respectively, both
at 95\% confidence level. This new analysis improves the limit on
the expected signal strength by almost a factor two assuming no $HH$
production compared to the previous iteration based only on 13\,\TeV
data~\cite{HDBS-2021-10}.

\FloatBarrier


%
%
%
%
%

\section*{Acknowledgements}

%

%

%
%

%
%

We thank CERN for the very successful operation of the LHC and its injectors, as well as the support staff at
CERN and at our institutions worldwide without whom ATLAS could not be operated efficiently.

The crucial computing support from all WLCG partners is acknowledged gratefully, in particular from CERN, the ATLAS Tier-1 facilities at TRIUMF/SFU (Canada), NDGF (Denmark, Norway, Sweden), CC-IN2P3 (France), KIT/GridKA (Germany), INFN-CNAF (Italy), NL-T1 (Netherlands), PIC (Spain), RAL (UK) and BNL (USA), the Tier-2 facilities worldwide and large non-WLCG resource providers. Major contributors of computing resources are listed in Ref.~\cite{ATL-SOFT-PUB-2026-001}.

We gratefully acknowledge the support of ANPCyT, Argentina; YerPhI, Armenia; ARC, Australia; BMWFW and FWF, Austria; ANAS, Azerbaijan; CNPq and FAPESP, Brazil; NSERC, NRC and CFI, Canada; CERN; ANID, Chile; CAS, MOST and NSFC, China; Minciencias, Colombia; MEYS CR, Czech Republic; DNRF and DNSRC, Denmark; IN2P3-CNRS and CEA-DRF/IRFU, France; SRNSFG, Georgia; BMFTR, HGF and MPG, Germany; GSRI, Greece; RGC and Hong Kong SAR, China; ICHEP and Academy of Sciences and Humanities, Israel; INFN, Italy; MEXT and JSPS, Japan; CNRST, Morocco; NWO, Netherlands; RCN, Norway; MNiSW, Poland; FCT, Portugal; MNE/IFA, Romania; MSTDI, Serbia; MSSR, Slovakia; ARIS and MVZI, Slovenia; DSI/NRF, South Africa; MICIU/AEI, Spain; SRC and Wallenberg Foundation, Sweden; SERI, SNSF and Cantons of Bern and Geneva, Switzerland; NSTC, Taipei; TENMAK, T\"urkiye; STFC/UKRI, United Kingdom; DOE and NSF, United States of America.

Individual groups and members have received support from BCKDF, CANARIE, CRC and DRAC, Canada; CERN-CZ, FORTE and PRIMUS, Czech Republic; COST, ERC, ERDF, Horizon 2020, ICSC-NextGenerationEU and Marie Sk{\l}odowska-Curie Actions, European Union; Investissements d'Avenir Labex, Investissements d'Avenir Idex and ANR, France; DFG and AvH Foundation, Germany; Herakleitos, Thales and Aristeia programmes co-financed by EU-ESF and the Greek NSRF, Greece; BSF-NSF and MINERVA, Israel; NCN and NAWA, Poland; La Caixa Banking Foundation, CERCA Programme Generalitat de Catalunya and PROMETEO and GenT Programmes Generalitat Valenciana, Spain; G\"{o}ran Gustafssons Stiftelse, Sweden; The Royal Society and Leverhulme Trust, United Kingdom; United States of America.

In addition, individual members wish to acknowledge support from CERN: European Organization for Nuclear Research (CERN DOCT); Chile: Agencia Nacional de Investigaci\'on y Desarrollo (ANID FONDECYT reg. 1230987, FONDECYT 1230812, FONDECYT 1240864, Fondecyt 3240661, Fondecyt Regular 1240721); China: Chinese Ministry of Science and Technology (MOST-2023YFA1605700, MOST-2023YFA1609300), National Natural Science Foundation of China (NSFC - 12175119, NSFC 12275265); Czech Republic: Czech Science Foundation (GACR - 24-11373S), Ministry of Education Youth and Sports (ERC-CZ-LL2327, FORTE CZ.02.01.01/00/22\_008/0004632), PRIMUS Research Programme (PRIMUS/21/SCI/017); EU: H2020 European Research Council (ERC - 101002463); European Union: European Research Council (BARD No. 101116429, ERC - 948254, ERC 101089007), European Regional Development Fund (HE COFUND GA No.101081355, ERDF), Horizon 2020 Framework Programme (MUCCA - CHIST-ERA-19-XAI-00), European Union, Future Artificial Intelligence Research (FAIR-NextGenerationEU PE00000013), Italian Center for High Performance Computing, Big Data and Quantum Computing (ICSC, NextGenerationEU); France: Agence Nationale de la Recherche (ANR-21-CE31-0013, ANR-21-CE31-0022, ANR-22-EDIR-0002, ANR-24-CE31-0504-01); Germany: Baden-Württemberg Stiftung (BW Stiftung-Postdoc Eliteprogramme), Deutsche Forschungsgemeinschaft (DFG - 469666862, DFG - CR 312/5-2); China: Research Grants Council (GRF); Italy: Istituto Nazionale di Fisica Nucleare (ICSC, NextGenerationEU), Ministero dell'Università e della Ricerca (NextGenEU 153D23001490006 M4C2.1.1, NextGenEU I53D23000820006 M4C2.1.1, NextGenEU I53D23001490006 M4C2.1.1, SOE2024\_0000023); Japan: Japan Society for the Promotion of Science (JSPS KAKENHI JP22H01227, JSPS KAKENHI JP22H04944, JSPS KAKENHI JP22KK0227, JSPS KAKENHI JP24K23939, JSPS KAKENHI JP24KK0251, JSPS KAKENHI JP25H00650, JSPS KAKENHI JP25H01291, JSPS KAKENHI JP25K01023); Norway: Research Council of Norway (RCN-314472); Poland: Ministry of Science and Higher Education (IDUB AGH, POB8, D4 no 9722), Polish National Science Centre (NCN 2021/42/E/ST2/00350, NCN OPUS 2023/51/B/ST2/02507, NCN OPUS nr 2022/47/B/ST2/03059, NCN UMO-2019/34/E/ST2/00393, UMO-2022/47/O/ST2/00148, UMO-2023/49/B/ST2/04085, UMO-2023/51/B/ST2/00920, UMO-2024/53/N/ST2/00869); Portugal: Foundation for Science and Technology (FCT); Spain: Agencia de Gestión de Ayudas Universitarias y de Investigación (AGAUR - 2023 BP 00141), Generalitat Valenciana (ASFAE/2022/008), Ministry of Science and Innovation (MCIN \& NextGenEU PCI2022-135018-2, MICIN \& FEDER PID2021-125273NB, RYC2019-028510-I, RYC2020-030254-I, RYC2021-031273-I, RYC2022-038164-I), Ministerio de Ciencia, Innovación y Universidades/Agencia Estatal de Investigaci\'on (PID2022-142604OB-C22); Sweden: Carl Trygger Foundation (Carl Trygger Foundation CTS 22:2312), Swedish Research Council (Swedish Research Council 2023-04654, VR 2021-03651, VR 2022-03845, VR 2022-04683, VR 2023-03403, VR 2024-05451), Knut and Alice Wallenberg Foundation (KAW 2018.0458, KAW 2022.0358, KAW 2023.0366); Switzerland: Swiss National Science Foundation (SNSF - PCEFP2\_194658); United Kingdom: Royal Society (NIF-R1-231091); United States of America: U.S. Department of Energy (ECA DE-AC02-76SF00515), Neubauer Family Foundation.

%
%


%
%
%
%
%
%
%
%

%
%

%
%
%
\printbibliography
\clearpage
 
\begin{flushleft}
\hypersetup{urlcolor=black}
{\Large The ATLAS Collaboration}

\bigskip

\AtlasOrcid[0000-0002-6665-4934]{G.~Aad}$^\textrm{\scriptsize 104}$,
\AtlasOrcid[0000-0001-7616-1554]{E.~Aakvaag}$^\textrm{\scriptsize 17}$,
\AtlasOrcid[0000-0002-5888-2734]{B.~Abbott}$^\textrm{\scriptsize 123}$,
\AtlasOrcid[0000-0002-0287-5869]{S.~Abdelhameed}$^\textrm{\scriptsize 119a}$,
\AtlasOrcid[0000-0002-1002-1652]{K.~Abeling}$^\textrm{\scriptsize 55}$,
\AtlasOrcid[0000-0001-5763-2760]{N.J.~Abicht}$^\textrm{\scriptsize 49}$,
\AtlasOrcid[0000-0002-8496-9294]{S.H.~Abidi}$^\textrm{\scriptsize 30}$,
\AtlasOrcid[0009-0003-6578-220X]{M.~Aboelela}$^\textrm{\scriptsize 45}$,
\AtlasOrcid[0000-0002-9987-2292]{A.~Aboulhorma}$^\textrm{\scriptsize 36e}$,
\AtlasOrcid[0000-0001-5329-6640]{H.~Abramowicz}$^\textrm{\scriptsize 157}$
% ...
\bigskip
\\

% ...
$^{17}$Department for Physics and Technology, University of Bergen, Bergen; Norway.\\
% ...
$^{30}$Physics Department, Brookhaven National Laboratory, Upton NY; United States of America.\\
% ...
$^{36}$$^{(a)}$Facult\'e des Sciences Ain Chock, Universit\'e Hassan II de Casablanca;$^{(b)}$Facult\'{e} des Sciences, Universit\'{e} Ibn-Tofail, K\'{e}nitra;$^{(c)}$Facult\'e des Sciences Semlalia, Universit\'e Cadi Ayyad, LPHEA-Marrakech;$^{(d)}$LPMR, Facult\'e des Sciences, Universit\'e Mohamed Premier, Oujda;$^{(e)}$Facult\'e des sciences, Universit\'e Mohammed V, Rabat;$^{(f)}$Institute of Applied Physics, Mohammed VI Polytechnic University, Ben Guerir; Morocco.\\
% ...
$^{45}$Physics Department, Southern Methodist University, Dallas TX; United States of America.\\
% ...
$^{49}$Fakult\"{a}t Physik, Technische Universit{\"a}t Dortmund, Dortmund; Germany.\\
% ...
$^{55}$II. Physikalisches Institut, Georg-August-Universit\"{a}t G\"ottingen, G\"ottingen; Germany.\\
% ...
$^{104}$CPPM, Aix-Marseille Universit\'e, CNRS/IN2P3, Marseille; France.\\
% ...
$^{119}$$^{(a)}$New York University Abu Dhabi, Abu Dhabi;$^{(b)}$United Arab Emirates University, Al Ain; United Arab Emirates.\\
% ...
$^{123}$Homer L. Dodge Department of Physics and Astronomy, University of Oklahoma, Norman OK; United States of America.\\
% ...
$^{157}$Raymond and Beverly Sackler School of Physics and Astronomy, Tel Aviv University, Tel Aviv; Israel.\\
% ...
$^{a}$ Also at Affiliated with an institute formerly covered by a cooperation agreement with CERN.\\
$^{b}$ Also at An-Najah National University, Nablus; Palestine.\\
$^{c}$ Also at Borough of Manhattan Community College, City University of New York, New York NY; United States of America.\\
$^{d}$ Also at Center for Interdisciplinary Research and Innovation (CIRI-AUTH), Thessaloniki; Greece.\\
$^{e}$ Also at Centre of Physics of the Universities of Minho and Porto (CF-UM-UP); Portugal.\\
$^{f}$ Also at CERN, Geneva; Switzerland.\\
% ...

\end{flushleft}


%

%
%
\clearpage

\end{document}